\documentclass{aa}
\usepackage{epsfig}
\usepackage{rotating}

\newcommand{\lesssim}{\mathrel{\hbox{\rlap{\hbox{\lower4pt\hbox{$\sim$}}}\hbox{$<$}}}}
\newcommand{\gesssim}{\mathrel{\hbox{\rlap{\hbox{\lower4pt\hbox{$\sim$}}}\hbox{$>$}}}}

\newcommand{\bmv}{$B-V$}

\newcommand{\teff}{$T_{\rm eff}$}
\newcommand{\nli}{$\log n$(Li)}

\begin{document}

\title{Time scales of Li evolution: a homogeneous analysis
of open clusters from ZAMS to late-MS}

   \subtitle{ }

   \author{P. Sestito\inst{1} \and S. Randich\inst{2}}

   \offprints{P. Sestito, email:paola.sestito@bo.astro.it}

\institute{INAF/Osservatorio Astronomico di Bologna, Via C. Ranzani 1,
            I-40127 Bologna, Italy
\and
INAF/Osservatorio Astrofisico di Arcetri, Largo E. Fermi 5,
             I-50125 Firenze, Italy}

\titlerunning{Li in open clusters}
\date{Received Date: Accepted Date}

\abstract{
We have performed a new and homogeneous analysis of all the Li
data available in the literature
for main sequence stars (spectral-types
from late F to K) in open clusters. 
In the present paper we focus on a detailed investigation
of MS Li depletion and its time scales for stars in the 6350--5500~K
effective temperature range.
For the first time, we were able to constrain the age at which
non-standard mixing processes, driving MS Li depletion, appear.
We have also shown that MS Li depletion is not a continuous
process and cannot be simply described by a $t^{-\alpha}$ law.
We confirm that depletion becomes ineffective beyond an age of 1--2 Gyr for the
majority of the stars, leading to a Li plateau at old ages.
We compared the empirical scenario of Li  as a function of age
with the predictions of three non-standard models. We found that
models including only gravity waves as main mixing process
are not able to fit the Li vs. age pattern
and thus this kind of mixing
can be excluded as the predominant mechanism responsible
for Li depletion. On the other hand, models including slow mixing 
induced by rotation and
angular momentum loss, and in particular those including
also diffusive processes not related to rotation,
can explain to some extent the empirical evidence.
However, none of the currently proposed models can fit the plateau
at old ages.
\keywords{ Stars: abundances -- Li --
           Stars: Evolution --
           Open Clusters and Associations: Individual: }}
\maketitle
\section{Introduction}\label{intro_lit}

Lithium, as well as the other light elements beryllium
and boron, is burned at relatively
low temperatures in stellar interiors
($\sim2.5\times10^{6}$ K). As a consequence, it
survives only in the external layers of a star
and it is a very powerful tracer
of mixing mechanisms at work in stellar structures.

A huge amount of
observational and theoretical work
has been devoted to the understanding of Li and
its evolution.
In particular,
during the last two decades, several datasets for
Li in open clusters and in both
Population {\sc i} and {\sc ii} field stars have been collected.
Focusing on Pop.~{\sc i},
observations of unevolved stars in open clusters and the field evidenced
several features that disagree
with the predictions of the standard models. With
``standard'' or ``classical''
we refer to those models that include convection only as a mixing
process and do not take into account transport phenomena
like diffusion, gravity waves, angular momentum loss and transport, etc.
The main open problems concerning Li evolution during the
pre-main sequence (PMS) and main sequence (MS) phases
of solar-type and lower mass stars have been recently discussed by
Jeffries (\cite{jef_cast04}) and Randich (\cite{R05_cast04}) and
are summarized below:

{\it (i)} PMS: G-type stars hotter than $\sim$ 5300--5400 K in 
zero-age main sequence ( ZAMS, $\sim$ 30--100 Myr) open clusters
(e.g., Soderblom et al. \cite{S93b} --- hereafter S93b;
Randich et al. \cite{R98}; Randich et al. \cite{R01})
have Li abundances (\nli)
only slightly below the initial abundance 
for Pop.~{\sc i} stars (\nli$_{0}=3.1-3.3$, see, e.g., Jeffries
\cite{jef_cast04}); a larger amount of PMS Li depletion is instead expected
from standard models. Note however that
the predicted Li destruction strongly depends on the physical assumptions
adopted in stellar codes.
In addition, 
K-type stars of presumably similar effective temperatures (\teff)
in clusters younger than $\sim 250$~Myr 
are characterized by a wide dispersion in Li abundance
whose origin is not yet understood, although different explanations
have been proposed.

{\it (ii)} MS: at variance with standard model predictions,
the empirical evidence for clusters older than the \object{Pleiades}
(e.g., the \object{Hyades}; Thorburn et al. \cite{T93} ---hereafter T93)
suggests that solar-type stars do deplete their photospheric
Li after having reached the ZAMS.
Another puzzling feature concerns the discovery of a ``Li dip''
for F-type stars around $\sim$6700--7000 K (e.g.
Balachandran \cite{bala95} and references therein):
stars of the Hyades and clusters of similar (and older) ages
within this \teff~range have Li abundances lower by a factor
$\sim$ 30 (or more) 
than those of stars within a \teff~range of $\sim\pm$ 300 K.
Such a gap in the Li distribution may start appearing at ages
as young as $\sim$200 Myr
(see Steinhauer \& Deliyannis \cite{SD04} and references therein).

Observations of old open clusters show that past the Hyades age Li
depletion for solar analogs can be either very slow or very fast:
in fact, solar-type
stars in the solar age/solar metallicity \object{M~67} are characterized by
a large amount of dispersion in
Li (see Pasquini et al. \cite{P97}, Jones
et al. \cite{jones99} and references therein), with about 40 \% of
the stars more Li depleted (by factors 5--10) than the remaining
fraction (``upper envelope''). On the other hand, the other old
open clusters investigated so far (\object{IC~4651}
and \object{NGC~3680}, Randich et al. \cite{R00};  \object{NGC~188},
Randich et al. \cite{R03}; NGC~752, Sestito et al. \cite{sestito04}) have
tight Li distributions with abundances similar to those in the
upper envelope of M~67 and only $\sim$2 times lower than those of
similar stars in the Hyades. We mention that preliminary analysis
of the old cluster \object{Cr 261} suggests that this cluster could
also be characterized by a dispersion, although not as large as
the one observed in M~67 (Pallavicini et al. \cite{pal05}).
A large spread similar to that present in M~67
is also observed for solar analogs in the field (e.g., Pasquini
et al.~\cite{P94}); the Sun is representative
of Li-poor stars in the field, and its
Li abundance cannot be reproduced by standard models.

{\it (iii)} Finally, Li depletion does not seem to significantly depend on the
cluster metal content; both young and old clusters with different
metallicities --- varying within
$\sim\pm0.2$ dex around the solar
[Fe/H] --- seem to share a very similar Li distribution
(e.g., Jeffries \& James
\cite{blanco1}; Sestito et al.~\cite{sestito03}, \cite{sestito04}). 

These puzzling
features, not expected from standard models,
suggest that Li depletion is not driven
uniquely by convection, and that extra-mixing processes
(or processes able to inhibit Li depletion during the PMS)
not included in classical theory
are at work in stars during the various evolutionary phases.
Several efforts have been done in the last years
aimed to understand this (these) process(es) and non-standard models have been
developed, but so far the mechanism(s) driving Li depletion
remain(s) poorly constrained.

In the present paper we focus on point {\it (ii)} above, 
and more specifically on the detailed
investigation of
MS Li depletion and its time scales in solar-type stars.
Since proposed non-standard mixing processes
have different time scales of Li depletion, our main goal is to
put empirical constraints on these time scales for stars with different
temperatures/masses and thus to provide feedback to the models.
In order to carry out such an analysis,
not only
the number of observed open clusters and of stars per cluster
should be enlarged to have a good age sampling, but --- most important ---
all the already available
Li data for open clusters should be analyzed
with the same method, to reduce the possibility
of spurious results due to an inhomogeneous analysis.

For this reason, we collected all the available information
on Li in late F, G and K-type MS stars in open clusters
and we performed a new analysis of Li abundances\footnote{The database is available upon request to P.~Sestito.}.
Our final objective is
to provide a homogeneous (at least as far as the analysis is concerned)
database for open clusters, that can
be used to study Li evolution and to test theoretical models.
We stress that, whereas most of the papers on Li in 
open clusters have so far focused on
the comparison of samples with similar ages or in a restricted age
range, our goal here is to investigate the evolution of Li and its
timescales from the ZAMS up to very old ages.
The database
will be also made available in the web.
Note that the discussion in the present paper focuses
on MS Li depletion in solar-type stars, but we included
in the database also warmer and cooler stars (F and K-type) in order to provide
a more complete picture for Li evolution in open clusters.
F and K-type stars will not be discussed here, since they are characterized by
different structures and evolutionary histories with respect to solar analogs
and deserve to be studied separately.

As a final but fundamental remark we would like to emphasize that,
even though our method of analysis can be affected by systematic errors,
we are interested in the investigation
of differential depletion of Li during the MS
and not in the absolute abundance values for each cluster: in this context,
the importance of performing a consistent analysis with all
the samples on the same scale appears even more evident.

In Sect.~2 we 
present the data sample, while in Sect.~3 the method
of analysis is described; Sect.~4 includes
the results and a discussion; 
some conclusions are reported in Sect.~5.

\setcounter{table}{0}
\begin{table*}[!] \scriptsize
\caption{\small Sample clusters and their parameters.}\label{samples_lit}
\begin{tabular}{ccccccccccccc}
\hline
\hline
Cluster & Ref. &Age &Ref. &$m_{v}-M_{v}$& $E(B-V)$ & Ref. &[Fe/H] & Ref. &\teff~range & No. of & $S/N$\\
        &Li data &(Gyr) & age & (mag) &(mag) &distance, reddening  &            &   [Fe/H]&(K) &stars (*) &\\
\hline
NGC~2264 & 1 &0.005&2 &9.4 &0.071 & 57 &$-$0.15&77&4050--6350 &12(4) & 50--100\\
          &2&& & & & && &4100--5550&37(8)&25--105 \\

\object{IC~2602}&3,4 &0.030&43 &6.11 &0.04 &58 &$-$0.05&4 &4150--6450 &31(1) &30--100\\

IC~2391&5 &0.030&5 &6.05 &0.01 &59,60 &$-$0.03&4 &4100--6050 &6&--\\

\object{IC~4665}& 6& 0.035&44 &8.3 &0.18 &61 &-- &--& 4750--5950&13 &--\\

NGC~2547 &7 &0.035&45 &8.30 &0.06 & 62&sub-solar& 7 &4100--5100& 7(2) &80--190\\
$\alpha$ Per&8,9 &0.050&8 &6.06 &0.08 &63 &$-$0.05&78 &4450--6450 &39 & 100--200\\
 &10 &&& & & & & &4150--5400 &18&--\\

\object{NGC~2451}&11 &0.050&46 &7.57 &0.05 &64,65 &$-$0.01&11 &4350--6350 &17(5) & 20--150\\

Pleiades&12 &0.07&47 &5.36 &0.04 &66,67 &$-$0.03&78 &4100--4700 &7 &--\\
 &13& & && & & & &4000--6500 &95(22)&--\\
 &14& & && & & & &4000--4250 &4&90--300\\
 &15& & && & & & &4050--5250 &10&$\sim$70\\
 &16& & && & & & &4750--5150 &9(1) &90--110\\

\object{Blanco 1}&17 &0.1&48 &6.9 & 0.02&68 &+0.14&17 &4550--5850 &17 &$\sim$30\\

\object{NGC~2516}&18 &0.15&49 &8.18 &0.12 &69 &$\sim-0.10$ &18 &4810--6250 &22(3) &10--65\\

\object{M~35}&19 &0.2&50 &9.60 & 0.225&50 &$-$0.21 & 19 &4750--6250 &27& 40--160\\

\object{NGC~6475}&20 &0.25&49 &7.35 &0.06 &70,20 &+0.14 &22 &4450--6400 &24(6)& 15--50\\
&21& && && & & &5150--6200 &10&--\\
&22& && && & & &5000--6300 &32(11) &50--150\\

\object{M~34}&23&0.25&49 &8.28 &0.07 &71,72 &+0.07 &79&4150--6450 &38(3)&$\sim$70\\

\object{Coma Ber}&24&0.6&51 &4.71 &0.013 &73,64 &$-$0.05 &80 &6200 &1(1) &75--105\\
&25& && && & & &4200--6300 &14(4)&150--250\\
&26& && && & & &4950--6200 &13(2)&200\\

\object{NGC~6633}&27&0.6&52 &7.77 &0.017 &27 &$-$0.10 &27 &4600--6500 &22(3)&$\sim$100\\

Hyades&24,28& 0.6&53 &3.31 &0.01 &53,64&+0.13 &78 & 4900--6200&14(1)&90--120\\
&29& && && & & &4600--6400 &58(31)&150--220\\

\object{Praesepe}&30&0.6&30 &6.39 &0.009 &64&+0.03 &80 &4850--6450 &57(22) &$\sim$100\\

NGC~752&31&2&54 &8.25 &0.035 &74 &+0.01 &32 &4900--6350 &5&30--80\\
&32&& & && & & &5550--6300 &18(1) &50--80\\

NGC~3680&33&2&54 &10.5 &0.05 &75 &$-$0.17 &34 &5900--5950 &2(1) &30--60\\
& 34&& &&& & & &5900--5950 &4(1) &60--150\\

IC~4651&33&2&54 &10.1 &0.083 &76 &+0.10 &81 &5800--6250 &10(1)& 30--60\\

M~67&35& 5&55 &9.69 &0.05 &56,39&$+$0.05 &82 &5650--6200 &6& --\\
&36& & & && & & &5900--6000 &3 &20--70\\
&37& & & && & & &6150--6200 &3(1)& 50--100\\
&38& & & && & & &5900--6150 &7(2)& $\sim$50\\
&39& & & && & & &5500--6150 &25(1)& 60--100\\
&40& & & && & & &5900--6150 &2& 100--200\\

NGC~188 &41&8&56 &11.44 &0.09 &56 &+0.01 &42&5750--5950 &2& $\sim$45\\
&42& && && & & &5700--6000 &11&20--35\\
\hline
\end{tabular}
* = The number in brackets in Col.~11 indicates the number of binaries
present in the samples. 

References: (1) King  (\cite{king}); (2) Soderblom et
al.~(\cite{sod2264}); (3) Randich et al.~(\cite{R97}); (4)
Randich et al.~(\cite{R01}); (5) Stauffer
(\cite{stauffer2391}); (6) Mart\'{\i}n \& Montes
(\cite{mm97}); (7) Jeffries et al.~(\cite{jef2547}); (8)
Balachandran et al.~(\cite{bala88}); (9) Balachandran et
al.~(\cite{bala96}); (10) Randich et al.~(\cite{R98});
(11) H\"unsch et al.~(\cite{hun04});
(12) Butler et al.~(\cite{butler});
(13) Soderblom et al.~(\cite{S93b}; S93b); (14) Garc\'{\i}a
L\'opez et al.~(\cite{GL94}); (15) Jones et
al.~(\cite{jones96}); (16) Jeffries (\cite{jefple}); (17)
Jeffries \& James (\cite{blanco1}); (18) Jeffries et
al.~(\cite{jef2516}); (19) Barrado et
al.~(\cite{barrado35}); (20) James \& Jeffries
(\cite{JJ97}); (21) James et al.~(\cite{J00});
(22) Sestito et al.~(\cite{sestito03}); (23) Jones et
al.~(\cite{jones97}); (24) Soderblom et al.~(\cite{S90});
(25) Jeffries (\cite{jefcoma}); (26) Ford et
al.~(\cite{fordcoma}); (27) Jeffries et
al.~(\cite{jef6633}); (28) Soderblom et al.~(\cite{S95});
(29) Thorburn et al.~(\cite{T93}; T93); (30) Soderblom et
al.~(\cite{S93c}); (31)  Hobbs \& Pilachowski
(\cite{HP86}); (32) Sestito et
al.~(\cite{sestito04}); (33) Randich et al.~(\cite{R00}); (34) Pasquini et al.~(\cite{P01}); (35)  Hobbs \&
Pilachowski (\cite{HP86b}); (36) Spite et
al.~(\cite{spite87}); (37) Garc\'{\i}a L\'opez et
al.~(\cite{GL88}); (38) Pasquini et al.~(\cite{P97}); (39)
Jones et al.~(\cite{jones99}); (40) Randich et al.~(\cite{R02});
(41)Hobbs \& Pilachowski
(\cite{HP88}); (42) Randich et al.~(\cite{R03});
(43) Stauffer et al.~(\cite{stauffer97}); (44) Mermilliod et al.~(\cite{mermi});
(45) Naylor et al.~(\cite{naylor}); (46) H\"unsch et al.~(\cite{hun03});
(47) Patenaude (\cite{pate});
(48) Panagi \& O'Dell (\cite{panagi}); (49) Meynet et al.~(\cite{meynet});
(50) Sung \& Bessel
(\cite{phot_sung99}); (51) Garc\'{\i}a L\'opez et al. (\cite{GL00});
(52) Strobel (\cite{strobel}); (53) Perryman et
al.~(\cite{phot_perryman}); (54) Friel (\cite{friel95});
(55) Hobbs \& Thorburn (\cite{ht91}); (56) Sarajedini et
al.~(\cite{Sa99})
(57) Sung et al.~(\cite{phot_sung97}); (58) Whiteoak
(\cite{phot_white});  (59) Becker \& Fenkart
~(\cite{phot_bf70});  (60) Patten \& Simon
~(\cite{phot_ps96}); (61) Prosser (\cite{phot_prosser93});
(62) Claria (\cite{phot_claria});  (63) Mitchell
(\cite{phot_mitch});  (64) Lyng\aa~Catalogue (1987 --- fifth
edition; Lyng\aa~\cite{lynga});
(65) Feinstein (\cite{fein});
(66) Robichon et
al.~(\cite{phot_robichon});  (67) Stauffer \& Hartmann
(\cite{phot_sh87});  (68) Epstein  \& Epstein
(\cite{phot_epep}); (69) Dachs \& Kabus
(\cite{phot_dk89});  (70) Snowden (\cite{phot_snowden});
(71) Ianna \& Schlemmer (\cite{phot_is93});  (72) Jones \&
Prosser (\cite{jp96});  (73) Odenkirchen et
al.~(\cite{phot_oden});  (74) Daniel et
al.~(\cite{daniel});  (75) Nordstr\"om et
al.~(\cite{phot_nord}); (76) Anthony-Twarog et
al.~(\cite{phot_AT});  (77) King (\cite{king93});  (78)
Boesgaard \& Friel (\cite{bf90});  (79) Schuler et
al.~(\cite{schuler});  (80) Friel \& Boesgaard
(\cite{fb92});  (81) Pasquini et al.~(\cite{P04});  (82)
Randich et al.~(\cite{R05}).
\end{table*}

\section{Data samples}\label{lit_data}

Our sample includes 22 clusters
(listed in Col.~1 of Table~\ref{samples_lit}) with
different ages and metallicities (Cols.~3 and 8; the
references for age and [Fe/H] are reported in Cols.~4 and 9 respectively, while
Col.~2 lists references for Li data).
Cols. 5 to 7 list the distance modulus
and reddening with the proper reference(s).
For the present analysis we considered stars
with 4000 $\leq$ \teff $\leq$ 6500 K; as mentioned,
the discussion in Sect.~4 will concern
solar-type stars, i.e. objects with 5500 $\leq$ \teff $\leq$ 6350 K.

The ages of the clusters
 range from $\sim$ 5 Myr for \object{NGC~2264}, to $\sim$ 6--8 Gyr for
NGC~188, while [Fe/H] values are within an interval of
$\sim\pm$0.2 dex from solar.
We considered the ``classical'' ages for each cluster,
i.e. those determined from isochrone fitting.
We would like to stress here that, as Table~1 shows,
our assumptions for cluster ages are a collection
of independent determinations, i.e. they were derived by different
authors and using disparate models. Thus our dataset is not homogeneous
as far as ages are concerned. In addition, we mention that for four
young clusters included in our sample
(\object{IC~2391}, \object{NGC~2547}, \object{$\alpha$ Per}, Pleiades)
new independent estimates of the ages have been obtained
based on the position of the ``Li depletion boundary'' (LDB;
see Stauffer et al.~\cite{LDBple}, \cite{LDBaPer}; Barrado
y Navascu\'es et al.~\cite{LDB2391}; Jeffries \& Oliveira~\cite{JO05}).
LDB ages are in general higher than the classical ones:
in the references mentioned above ages of 50 Myr
for IC~2391, 90 Myr for $\alpha$ Per and 125 Myr for the Pleiades have been
reported, to be compared with those listed in Table~1.
In the case of NGC~2547 (Jeffries \& Oliveira~\cite{JO05}) there is
instead a perfect agreement between the LDB age and the classical one ($\sim$35 Myr).

The classical methods for estimating the age of a cluster,
i.e. the fit of the turn off or of the low MS,
are affected by uncertainties mainly due to physical
inputs adopted in models; on the contrary, the LDB
technique is less model dependent. However,
we adopt here the classical ages to be consistent with
the whole sample, since LDB
ages are available only for four clusters. 

The metallicities listed in Table~\ref{samples_lit} were all derived
from high-resolution data (see the quoted references), with
the exception of those of NGC~2264 and NGC~6633 that
were estimated from intermediate resolution spectra
(King \cite{king93};
Jeffries et al.~\cite{jef6633});
however these are the most recent
and reliable estimates for the Fe content of the two samples.
No estimate of the metallicity is available for IC~4665;
for NGC~2547 we only indicate that the cluster has a sub-solar Fe
content (see also Jeffries et al.~\cite{jef2547}) since no precise estimate
is present in the literature. As in the case of ages, we {\em caveat} that
also the [Fe/H] values for most clusters were not derived with a homogeneous
analysis, with exception of IC 2391, IC 2602, NGC~2451,
NGC 6475, NGC 752, M67 for
which we have derived the metallicity on the same scale as for the Hyades.

Several clusters were studied by more than one author; in these cases
we identified the stars in common between the different samples and we
selected the more recent observations (which have in most cases a better
signal-to-noise ratio --- $S/N$ --- and/or higher
resolving powers --- $R$); among the various clusters
the data selection for the Pleiades deserves a slightly more accurate
description.
For this cluster we adopted the following priority order:
(1) data of Jeffries (\cite{jefple}); the author carried out observations
of a sample of cool Pleiades stars
in order to investigate
the variability of Li~{\sc i} and K~{\sc i} and its possible relationship
with the spread among K-type objects.
For this reason
several objects were observed twice
(November 1997
and November 1998): since the equivalent widths ($EW$) measured from
the two sets of spectra are in good agreement
(no Li variability was found over a 1 year time scale)
we chose the value obtained from
the spectrum with the highest $S/N$ ratio.
In one case the two spectra have the same $S/N$
and thus
we computed an average of the $EW$s;
(2) data of Jones et al. (\cite{jones96}; although
they have lower $S/N$ than those of Garc\'{\i}a L\'opez et al. (\cite{GL94}),
the resolution is higher);
(3) data of S93b,
if the spectra
has a ``a'' quality (see the quoted reference), otherwise (4)
data of Garc\'{\i}a L\'opez et al. (\cite{GL94}).
We preferred the ``a'' quality data of S93b with respect to those
of Garc\'{\i}a L\'opez et al. (\cite{GL94}) since they have in principle
a slightly better $S/N$; we note however that this subsample
contains only two stars, for which the errors in $EW$ quoted
by the two authors are very similar; (5) data of Butler
et al.~(\cite{butler}). Note that S93b
re--observed some stars included in the sample of Butler et al., 
but they quoted
a larger error in $EW$ with respect to
the previous paper (and thus, probably a worse $S/N$). 
For this reason, we retained the original measurements by 
Butler et al.~(\cite{butler}). Finally, (6) all the remaining
stars included in the sample of S93b.

The number of stars and the relative range of effective
temperatures (as computed by us, see
Sect.~\ref{teff_lit}) for each sample are reported in Cols.~11 and 10 of
Table~\ref{samples_lit}, while the references
for the source papers are listed in Col.~2.
Most obviously only confirmed cluster members were considered. The
number in brackets in Col.~11 represents the number of binaries
present in the samples; double systems can be safely included in
our analysis since in general they do not affect the Li
distributions (but see also Sect.~\ref{timescales_lit}).

We finally mention that the cluster datasets
considered here were observed
during the last $\sim$ 20 years using different telescopes
and/or instruments; therefore, the various subsamples
are characterized by disparate $S/N$
and $R$.
Spectral resolutions range from $\sim$ 1 \AA~in the case of
NGC~2264 (Soderblom et al.~\cite{sod2264}) up to $\sim$ 0.1~\AA~($R\sim57,000$) for
NGC~752 and NGC~188 (Sestito et al.~\cite{sestito04};
Randich et al.~\cite{R03});
$S/N$ ratios (listed
in Col.~12
of Table~\ref{samples_lit})
range approximatively from $\sim$ 20--30 --- e.g., in the cases
of Blanco~1 (Jeffries \& James \cite{blanco1}) and
of the old and rather distant cluster NGC~188 (Randich et
al.~\cite{R03}) --- up to the very good quality
of the spectra of some stars in in the Pleiades 
($S/N\sim200-300$ --- Garc\'{\i}a L\'opez et al. \cite{GL94}),
in $\alpha$ Per ($S/N\sim100-200$ --- Balachandran et al.~\cite{bala88}),
in NGC~6475 ($S/N\sim150$ ---
Sestito et al.~\cite{sestito03})
and other young/intermediate age clusters.
In other words, from the point of view
of the data quality the complete set cannot be considered homogeneous;
homogeneity would imply that all the stars in all the clusters
should have been re-observed with a unique instrument
and possibly in similar weather conditions. 
Thus, a homogeneous dataset on the point
of view of $S/N$ cannot be obtained,
since also spectra collected during the same night
can be characterized by discrepant qualities due to
changes in weather conditions.

Errors in $EW$s depend on the $S/N$ ratio of the spectra, but
also
the spectral resolution affects the estimate
of the Li $EW$s: as we will explain in more detail
in Sect.~\ref{Li_lit} this spectral feature can be blended
with a nearby Fe~{\sc i} line. When the resolution is not large enough
(or in presence of large rotation) the two lines cannot be separated and
we subtract the contribution of the Fe~{\sc i} feature
to the total $EW$ using an analytical expression
in which $EW$(Fe) is a function of $B-V$.
As a consequence, Li abundances derived from spectra
with different resolving powers are differently affected by
errors introduced with the
theoretical estimate of the Fe contribution (see Sect.\ref{random_lit}).

\section{Li analysis}\label{analysis_lit}

\subsection{Effective temperature}\label{teff_lit}

We started our new analysis from the published $EW$s
of the Li~{\sc i} $\lambda\rm{6707.79\,\AA}$ line and from $BV$
photometry
used in the original papers;
the proper references will be
reported in the web database.
As already mentioned,
the adopted reddening values towards each cluster
are listed in Col.~6 together with the reference (Col.~7).
Note that the $E(B-V)$ values adopted for each cluster are
rather ``safe'', in the sense that discrepancies
with other estimates reported in the
literature are very low ($\sim$ 0.03 mag at most);
this is due to the fact that the clusters considered in the current analysis
are relatively close and well studied. In addition, none of the sample clusters
is characterized by differential reddening (see the database by
J.-C.~Mermilliod and references therein --- http://obswww.unige.ch/webda).

We decided to use $BV$ photometry
since for the large majority of F, G and K-type stars
there are by far more \bmv~observations
available than for any other color index:
this allows
us not to restrict the dataset too much
(for example $VI$ photometry is in general available
only for cool stars).
In addition, and most important, it has been shown
(Soderblom et al.~\cite{S93a} --- hereafter S93a; T93)
that \bmv~colors provide reliable \teff~values at least for F and
G-type stars, the latter ones being the objects of main interest
for our investigations; in fact, photometric temperatures obtained
from $B-V$ color indices are in general
in good agreement with temperatures obtained
via spectroscopic analysis (see, e.g., Sestito et al.~\cite{sestito04}).
On the other hand, \bmv~are not very good
\teff~indicators for cool and/or active stars: first of all there is
an intrinsic effect due to the fact that
at lower \teff~the peak of the blackbody distribution
shifts towards lower energies (higher wavelengths);
for this reason, the dependence
of the effective temperature on $B-V$ becomes rather flat
for $B-V$ above $\sim$ 1 (see, e.g., Gray \cite{gray};
Bessel \cite{bessel}).
Second, as discussed by S93a,
in ultra-fast rotators, which represent a significant
fraction of young late-type objects, $B-V$ colors can be
altered either by effects of surface activity (e.g., spots;
Stauffer et al.~\cite{stauffer84}) or the color
distortion may arise from structural changes due to rotation.

In order to check on possible differences among
\teff~derived from different color indices, particularly for young cool
and active stars, we show in
Fig.~\ref{lit_Teff_cal_VI} a comparison between the
\teff~derived by us from $BV$ photometry (see below) and those
published in the original papers for the young cluster IC~2602
(Randich et al.~\cite{R97}; \cite{R01}),  derived as an average of
the results from two calibrations (based on $B-V$ and $V-I$).
The figure indicates that the two sets of temperatures are in good agreement,
not only for warm stars, but also for cooler objects, implying that
the use of $B-V$ colors should not introduce major internal
errors in the present study.

Effective temperatures were computed by us
from dereddened \bmv~colors using the calibration of S93a:
$T_{\rm eff}=1808{(B-V)_{0}}^{2}-6103(B-V)_{0}+8899$ K;
the authors found this temperature scale to be reasonably consistent
with the temperatures used for the Pleiades 
by Boesgaard \& Friel (\cite{bf90}), who
performed a calibration using F-type stars, and
 with
that of Arribas \& Rogers (\cite{arribas}) derived from cool stars with
the IR flux method.

\subsection{Li abundances}\label{Li_lit}

The Li~{\sc i} feature can be blended
with a nearby Fe~{\sc i} line at $\rm{6707.44\,\AA}$,
depending on the spectral resolution; for this reason,
if necessary,
the $EW$s were corrected from the Fe contribution
following the prescriptions
of S93b:
$EW$(Fe)=20$(B-V)_{0}-3\,{\rm m\AA}$.
For the following samples
the resolving power was high enough to allow the separation
of the two features:
Pleiades of Jones et al. (\cite{jones96}), M~34 (Jones et al. \cite{jones99}),
NGC~752 (Sestito et al. \cite{sestito04}, Hobbs \& Pilachowski \cite{HP86})
 and
NGC~188 of Randich et al. (\cite{R03}).
On the other hand, James \& Jeffries (\cite{JJ97}) and
James et al. (\cite{J00}; for NGC~6475), Jeffries
\& James (\cite{blanco1}; Blanco~1) and Balachandran et al. (\cite{bala88},
\cite{bala96}; $\alpha$ Per) published $EW$s
already corrected from the Fe~{\sc i} contribution but with a method
different from that of S93b.
Sestito et al. (\cite{sestito03}) carried out a Li analysis of NGC~6475
and showed that, for stars in common
with James \& Jeffries (\cite{JJ97}) and
James et al. (\cite{J00}), there
is a good agreement between the different sets of deblended $EW$s.
Also Randich et al. (\cite{R98}), in the analysis of $\alpha$ Per
trusted in the Fe~{\sc i}~correction performed by Balachandran et al. (\cite{bala88}, \cite{bala96}).
Thus, we could safely use the deblended equivalent widths
reported by these authors.
We finally mention that
upper limits in $EW$ were never corrected for the Fe~{\sc i} blending.

Li abundances were computed from \teff~and deblended
$EW$s by an interpolation
of the curves of growth (COG) of S93b.
These authors tested the accuracy of the results obtained
from the COGs and reduced the possible systematic errors
by calculating also synthetic spectra for broad-line stars.
The two techniques were found to be in good agreement, while discrepant
results for a subsample of stars were attributed to the saturation of the
lines; a brief discussion of systematic errors
due to different methods of analysis is reported in Sect.~\ref{syst_lit}

The COGs of S93b are based on the assumption of local thermodynamic
equilibrium (LTE), thus Li abundances were corrected for non-local
thermodynamic equilibrium
($N$LTE) effects --- which are very important especially
in the case of cool stars --- using the prescription of
Carlsson et al. (\cite{carlsson}).

\subsection{Uncertainties in Li abundances}\label{errors_lit}
\subsubsection{Random errors}\label{random_lit}

Random errors in Li abundances derive mainly from
uncertainties in effective temperatures --- due
to photometry uncertainties --- and in $EW$s,
while other parameters, such as surface gravities and microturbulence
velocities do not significantly affect the determination of
Li abundance from the Li~{\sc i}~feature.

Random 
errors in $EW$s come obviously from uncertainties in the measurement,
depending on the $S/N$ of the spectrum, and in addition,
as mentioned in
Sect.~\ref{lit_data},
they depend on the spectral resolution, i.e.
on whether or not the Li~{\sc i}~doublet can be
resolved from the nearby Fe~{\sc i}~line.
In order to check how large can be the error introduced in estimating
$EW$(Fe) via the correction of S93b, we measured
its value in spectra of stars with different \teff~and
for which we were able to separate the
two features. We found that for solar-type stars (\teff~$\sim$ 5700 K;
in NGC~752) the measured $EW$(Fe) range between $\sim$ 9 and 12 m\AA,
while one would obtain $EW$(Fe) $\sim$ 10 m\AA~using
the S93b formula; for hotter stars (\teff~$\sim$ 6200 K)
the observed $EW$(Fe) is 5--8 m\AA~(analytical: $\sim$ 7 m\AA),
while for cooler stars (in the Hyades) around 5000 K the analytical estimate 
gives
$\sim$ 14 m\AA, to be compared with
our measurement of $EW$(Fe) ranging from 13 up to 18.
Therefore, the largest difference between the measured $EW$s(Fe)
and those estimated with the S93a prescriptions
is of the order of 5 m\AA:
this difference can be significant only in
the worse case of rather cool stars (below 5000 K) and low $EW$
($\sim$ 10--20 m\AA) since it would 
correspond to an error in \nli~of $\sim$ 0.15 dex.
Note however that stars with \teff~below 5000 K in young cluster have
much larger Li $EW$s; on the other hand, the few measurements
at these temperatures in older clusters (e.g., Coma Ber)
are upper limits in Li, from which the Fe~{\sc i} contribution is not
subtracted.
On the contrary, for solar-type stars in old clusters
(e.g., with \teff~$\sim$ 6000 K and $EW$ $\sim$ 50 m\AA),
even in the most conservative case,
an error of 5 m\AA~would correspond to 0.05 dex in \nli, which
is well below the uncertainties usually estimated for these
objects.

Errors in $EW$s were published
in the original paper only for a part of
the clusters in the dataset; for these samples we retrieved
the published $\Delta{EW}$ and
we assumed
errors in effective temperatures according to the
uncertainties in the photometry or in the same
\teff, when quoted by the authors using \teff~calibrations
consistent with ours.
Thus, we were able to estimate
errors in Li abundances by quadratically adding $\Delta{EW}$ and
$\Delta$\teff.

On the other hand, for
about half of the sample (including the most populous ones --- Pleiades
of S93b and Jones et al. \cite{jones96},
Hyades of T93 and Praesepe) $\Delta{EW}$
 and $\Delta$\teff~for each star were not reported in the original
papers and the authors quoted only
typical average errors in these parameters or in \nli.
In those cases
we computed a mean error for each cluster starting
from the typical average uncertainties.

A detailed description of the error estimates for
each cluster is given in Appendix.

\subsubsection{Systematic errors}\label{syst_lit}

In order to investigate possible systematic effects on our temperature scale,
in Figure~\ref{lit_Teff_cal} we compare the \teff~values determined by us
 for the $\sim$ 100 Myr old open cluster Blanco~1 with those derived
by Jeffries \& James
(\cite{blanco1}): 
this is the sample for which we found the largest
difference between our \teff~scale and that used by previous
authors. Jeffries \& James (\cite{blanco1}) derived effective
temperatures for the hottest star ($B-V_{0}<0.64$) from $B-V$ colors
using the calibration of Saxner \& Hammarb\"ack (\cite{SH85}),
which contains a metallicity dependent term; for cooler objects
Jeffries \& James applied the B\"ohm-Vitense (\cite{bohm}) solar metallicity
calibration modified to take into account the higher than solar
Fe content of this cluster ([Fe/H] $=+0.14$;
Jeffries \& James \cite{blanco1}). 
Random errors in our \teff~were estimated from
uncertainties in the photometry (ranging from
$\Delta{(B-V)}\sim0.01$ up to 0.03), while for the temperatures of
Jeffries \& James we considered an average error bar similar to
those plotted in their Fig.~1 (since they do not quote a
$\Delta{T_{\mathrm{eff}}}$ value in their tables). A
reasonable agreement between the two samples is reached only for
the two hottest stars; in the other cases our temperatures are
systematically lower, with differences of $\sim$ 200 K for stars
around 5500 K, increasing up to almost 350 K for stars cooler than
5000 K.

In the case of other clusters analyzed by the group of Jeffries
(Coma Ber, NGC~2516, NGC~6633), all with
slightly sub-solar Fe contents, the differences between our
\teff~and those quoted in the original papers, derived using the
metal-dependent
Saxner \& Hammarb\"ack (\cite{SH85}) calibration, are rather
small and within the errors: this indicates that the
latter calibration is consistent with the S93a one, at least
for [Fe/H]$\lesssim$0.0. The large discrepancies
present in the case of Blanco~1 instead suggest that
when over-solar [Fe/H] are considered,
the metallicity correction for the \teff~vs.~\bmv~relationship
might be important. 

As a test, we derived effective temperatures for Blanco~1 using also
the metal dependent calibration by Alonso et al.~(\cite{alonso99})
and assuming [Fe/H]=+0.14. We found that for stars
with $B-V<0.70$ the \teff~derived with the Alonso
calibration are cooler than ours by $\sim$ 50 K.
On the contrary, for larger colors, the \teff~values
found using the Alonso calibration are hotter than ours, with differences
up to 250 K for the coolest star (\teff $\sim$ 4500 K).
This confirms that the [Fe/H] correction is not negligible
for over--solar metallicities and cool stars,
and it represents an additional warning for the future users
of the database. We stress however that
only four clusters in the database have
metallicities significantly
higher than solar (Blanco~1, NGC~6475, the Hyades and IC~4651) and,
as evidenced above, 
the \teff--($B-V$) relationship for solar-type stars 
--- the objects of main interest for our discussion in Sect.~4 ---
is not strongly affected by [Fe/H].

\begin{table*}[!] \footnotesize
\caption{Comparison between Li abundances derived
with the COGs of S93b and those derived with MOOG.}\label{MOOG}
\begin{tabular}{cccccccc}
\hline
\hline
Star & \teff & $EW$ & \nli & \nli  \\
     &  (K)  & m\AA &    MOOG        &   COGs (S93b)               \\
NGC~6475& & & & \\
R14 & 5888&  98   &  2.85&  2.76 \\
R39A&  5696&  98 &  2.64&  2.58 \\
R49A&  5772&  82 &  2.61&  2.54 \\
R102&  5079&  67 &  1.76&  1.69 \\
Hyades & & & & \\
vB22& 5477& 56& 2.12& 2.03\\
vB92& 5407& 15&1.36& 1.34\\
vB162& 5659& 59& 2.34& 2.25\\
NGC~752 & & & & \\
P475&5932 &59& 2.59& 2.50\\
P520&6151 &67& 2.86& 2.76\\
P983&5986 &45& 2.48& 2.40\\
\hline
\hline
\end{tabular}
\end{table*}

Systematic errors for the Li distributions of open clusters
are mainly due to the method of analysis;
we derive Li abundances with the COGs of S93b, but
slightly discrepant results would be obtained using a
different method.
In Table~\ref{MOOG} we show a comparison
between \nli~values obtained with the COGs of S93b (Col.~5)
and those obtained using the program MOOG (version 2002 --- Sneden
\cite{sneden};
Col.~4) for stars
with various $EW$s and \teff~in clusters of three different ages
(NGC~6475, the Hyades and NGC~752). 
In the computation with MOOG
we adopted the same surface gravity ($\log g=4.5$)
for all the stars, while the microturbulence was calculated
as $\xi=3.2 \times 10^{-4} (\rm T_{\rm eff}-6390)-1.3(\log g-4.16)+1.7$
(Nissen \cite{nissen},
Boesgaard \& Friel \cite{bf90}); we recall from Sect.~\ref{random_lit}
that Li abundances are not significantly affected by these two
stellar parameters.

Li abundances derived with MOOG and with the COGs appear to be
in reasonable agreement: whereas values derived with MOOG are
systematically higher than the others, all the differences are
$\leq$ 0.1 dex, i.e. smaller than (or
comparable with) typical errors quoted for the
three clusters (see Appendix);
note that 
the discrepancies in the final \nli~values decrease for
lower \teff~(in stars with similar $EW$).

\section{Results and discussion}\label{results_lit}
\subsection{New Li abundances}\label{new_lit}

Figure~\ref{lit_comparison} shows the comparison between the
``new'' Li abundances obtained with our method of analysis and
those reported in the original papers for two open clusters:
$\alpha$ Per (panel a) and Blanco~1 (panel b); filled symbols
represent our estimates. 
Error bars represent indicative average errors in \nli~and \teff;
in the case of $\alpha$ Per we found values
very similar to those published in the original reference (for this
reason we report only
one error bar), while for Blanco~1
our errors are slightly more conservative 
than those quoted by Jeffries \& James (\cite{blanco1}).

Data for $\alpha$ Per were taken from
Randich et al. (\cite{R98}); note that we included in the plot
also part of the sample of Balachandran et al.~(\cite{bala88};
\cite{bala96}) that was re-analyzed by Randich et al.
As evident, differences in the Li distributions do exist,
depending on the method of analysis. Figure~\ref{lit_comparison}a
shows that in the case of $\alpha$ Per the two sets of Li
abundances and \teff~are characterized by random differences;
however, in
most cases the differences in \teff~and \nli~are not very large
and they lie within the errors.

As shown in Fig.~\ref{lit_comparison}b and as already discussed,
in the case of Blanco~1
there are significant
systematic differences between our \teff~and those retrieved from the
original paper (see also Fig~\ref{lit_Teff_cal});
this example shows how systematic errors
due to different methods of analysis can affect the comparison
between open clusters.
Differences in Li
abundances  are only due to the discrepancies in \teff~since
also Jeffries \& James (\cite{blanco1}) derived
\nli~values using the COGs of S93b.
However, the two sets of \nli~appear to be in good agreement, and
the net effect is that the
two Li distributions are shifted one with respect to the
other. 

In summary
Fig.~\ref{lit_comparison} confirms that
 distinct methods of analysis result into random and systematic differences
in the
various sets of \teff~and \nli; therefore the sign and the size of the
error introduced by comparing clusters analyzed by different
authors cannot be estimated
\emph{a priori}. 
This reinforces the conclusion that
comparisons between 
clusters of different ages and with stars covering a wide range of
\teff~greatly benefit from the use of a unique method of analysis.

The complete dataset for Li in open clusters 
obtained with our
method of analysis 
is presented in Figs.~\ref{OC_YOUNG},~\ref{OC_IM} and~\ref{OC_OLD}:
in the first figure the Li distributions of young clusters
(ages ranging from few Myr up to $\sim$ 100 Myr) have been plotted,
while in the other two we show the distributions for intermediate 
age (150--600 Myr)
and old samples ($\gesssim$ 1 Gyr.) Binary stars
have been evidenced by filled symbols.
We will not describe in detail the Li vs. \teff~distributions, 
whose features have been widely discussed in the original papers;
we only note that the 
the star-to-star scatter  is indeed evident both for cool stars in
young clusters
and among solar-type stars in the M~67, 
and that binary stars in general
do not deviate from the average trends (see Sect.~\ref{timescales_lit}).

\subsection{Time scales of Li depletion}\label{timescales_lit}

In Fig.~\ref{lit_age_nonstd} we plot the average Li abundance as a
function of age for all the stars in the clusters in three
temperature ranges: $T_{\mathrm{eff}}=6200\pm150$ K (late F-type;
panel a), $T_{\mathrm{eff}}=5900\pm150$ K (G-type; b) and
$T_{\mathrm{eff}}=5600\pm100$ K (late G-type; c); we chose a
width of $\pm150$ K for the first two intervals in agreement with
the largest conservative uncertainties in \teff~for some clusters
(see Appendix); in the case of stars around 5600 K
we restricted instead the \teff~range since in clusters as old (or
older than) the Hyades the Li distributions are rather steep at
these temperatures.
Note that in some cases the cluster samples 
contain only few stars in the three \teff~ranges considered; thus,
in order to base our discussion on statistically more significant samples, the
datasets of clusters with similar ages have been merged. Specifically,
we merged the samples of IC~2602 with those of IC~2391 and
IC~4665 (age $\sim$ 30 Myr); $\alpha$ Per with NGC~2451
($\sim$ 50 Myr); Blanco~1 with the Pleiades
($\sim$ 100 Myr); M~34 with M~35 and NGC~6475 ($\sim$ 250 Myr);
the Hyades with NGC~6633, Coma Ber and Praesepe ($\sim$ 600 Myr); the three
clusters at 2 Gyr (NGC~752, NGC~3680 and IC~4651). Part of the
merged samples include clusters with different Fe contents:
however, as we have discussed in Sect.~\ref{intro_lit}
(and as visible from Figs.~\ref{OC_YOUNG}, ~\ref{OC_IM}
and~\ref{OC_OLD}),
observations of cluster with the same age, but disparate [Fe/H] clearly 
indicate that overall
metallicity does not affect Li depletion. For
this reason, we can safely merge the datasets of clusters with
similar age but different metallicity.
We add again some words of caution about this choice
to remind that ages and [Fe/H] values for the sample clusters are the results
of independent, and thus not homogeneous, investigations.
Error bars in Fig.~\ref{lit_age_nonstd} indicate 1$\sigma$ 
standard deviations from
the average; stars significantly deviating from the mean Li pattern
in each merged sample have not been included; in the case of M~67
we computed the mean \nli~separately for the upper envelope
and for the lower envelope. The average values for the lower
envelope in the 6200$\pm 150$~K
and 5900$\pm150$~K temperature ranges were computed using ASURV
that allows taking into account upper limits. For the coolest temperature
interval we can instead only provide an upper limit to the mean, since
we do not have enough data points.
The average \nli~values for the various clusters (or groups
of clusters) are summarized in Table~\ref{average}.

\begin{table*}[!] \scriptsize
\caption{\small Average \nli~in three ranges of \teff.}\label{average}
\begin{tabular}{cccccccc}
\hline
\hline
Cluster(s) & Age & \nli$\pm\sigma$ &\nli$\pm\sigma$&\nli$\pm\sigma$ \\
     &  (Gyr)  &  &  &   \\
\hline
& & $T_{\mathrm{eff}}=6200\pm150$ K  &$T_{\mathrm{eff}}=5900\pm150$ K  &$T_{\mathrm{eff}}=5600\pm100$ K \\
\hline
NGC~2264 &0.005 &3.34$\pm$0.10 &3.21$\pm$0.17 & 3.18 \\
IC~2391/IC~2602/IC~4665 &0.03 &-- &3.06$\pm$0.13 & 2.90$\pm$0.11 \\
$\alpha$ Per/NGC~2451 &0.05 &2.97$\pm$0.13 &2.96$\pm$0.10 & 2.94$\pm$0.09 \\
Pleiades/Blanco~1 &0.1 &2.95$\pm$0.09 &2.96$\pm$0.07 & 2.79$\pm$0.10 \\
NGC~2516  &0.15 &2.92 & 2.85$\pm$0.11 & 2.80$\pm$0.23\\
M~34/M~35/NGC~6475 &0.25 &2.92$\pm$0.13 & 2.79$\pm$0.14  & 2.57$\pm$0.19\\
Hyades/Praesepe/Coma Ber/NGC~6633 &0.6 & 2.77$\pm$0.21& 2.58$\pm$0.15  &2.14$\pm$0.22 \\
NGC~752/NGC~3680/IC~4651 & 2&2.65$\pm$0.13 & 2.33$\pm$0.17 & 1.91$\pm$0.24\\
M~67 -- upper envelope & 5&2.55$\pm$0.18 & 2.25$\pm$0.12&1.90$\pm$0.15 \\
M~67 -- lower envelope &5 &1.92$\pm0.05$ &1.32$\pm$0.09 & $\leq$1.29\\
NGC~188 &8 & -- &2.34$\pm$0.14 &2.30 \\
\hline
\hline
\end{tabular}
\end{table*}

With only few exceptions, binaries have been included in 
the determination of the average
abundance, since we find that their presence does not affect the Li
distributions, i.e.~their Li abundances in general follow the
average trend for each cluster (Figs.~\ref{OC_YOUNG}, \ref{OC_IM}
and~\ref{OC_OLD}; see also Sestito et al. \cite{sestito03}); 
for example, among the few binaries
excluded, there is
star S1045  in M~67
(with \teff~$\sim$ 6100 K) since, as discussed
in Pasquini et al.~(\cite{P97}), it lies significantly above
the mean Li trend. 
Also star RX76A (\teff~$\sim$ 4200 K)
 in NGC~2547 has an abnormally high Li abundance, and it is suspected
to be a tidally locked binary in which Li depletion was inhibited
(Jeffries et al.~\cite{jef2547}); anyway this is a very cool object, not included
in our discussion of the time scales of Li depletion.
We mention in passing that Barrado y Navascu\'es \&
Stauffer (\cite{barradohyades}) investigated a sample of
Hyades binaries already observed by T93 finding that, whereas the
general pattern is the same for double systems and single stars,
binaries show slight over-abundances, that appear more conspicuous
in tidally locked systems. Anyway, Barrado y Navascu\'es \&
Stauffer obtained higher $EW$s for binaries with respect to T93; in
our analysis we retain the $EW$ values published by T93 and we find
indeed that binary systems in the Hyades perfectly follow the
average Li pattern of the cluster.

The two dotted lines in the figure
represent the allowed range for the initial
Li abundance in Pop.~{\sc i} stars; the other curves represent
models including non-standard
physical processes (see below).
Standard models are not shown in the plots
since, as already mentioned, they do not predict
any MS Li depletion; on the contrary, we are interested
in the evolution of Li during this phase. As for the PMS phases, we 
recall that the amount of Li destroyed in this stage
is a strong function of the assumed input physics in the models;
different standard models result into distinct amount of PMS depletion
and thus into disparate \nli~values on the ZAMS
(e.g., D'Antona \& Mazzitelli \cite{dm94}; Siess et al.~\cite{siess}),
but, as stressed in Sect.~\ref{intro_lit}, our aim is the investigation of
the differential Li depletion during the MS, neglecting the
absolute ZAMS abundances of the stars.

Fig.~\ref{lit_age_nonstd}a shows that late F-type stars
($T_{\mathrm{eff}}=6200\pm150$ K) deplete a certain amount of Li
during the PMS phase; note that in this plot data for IC~2602,
IC~2391 and IC~4665 are lacking (since no observations are
available for stars within this \teff~interval), thus a lower
limit for Li abundances of ZAMS stars is represented by the mean
\nli~of $\alpha$ Per + NGC~2451 (age $\sim$ 50 Myr). Then, only a small
amount of destruction is suffered by these stars from 
50 Myr to $\sim$ 250 Myr; 
Li depletion slightly increases again beyond this age, as
suggested from the points representative of clusters at the age
of the Hyades and older
(NGC~752 + IC~4651 + NGC~3680 and the upper envelope of M~67);
Li destruction then appears to slow down
between 2 and 5 Gyr (see below --- plateau at old ages).
Finally, stars in the lower envelope of M~67 have 
depleted an amount of Li larger
by a factor $\sim$ 4 with respect to their Li-rich counterpart.

Panels b) and c) show that the Li evolution histories for stars
in the \teff~range 5500--6050 K are qualitatively similar to that
of hotter stars: these objects deplete Li by a factor lower than 2
(depending on the assumed initial content) during the PMS; then Li
depletion virtually slows down/stops between the age of $\alpha$
Per up to $\sim$ 150 Myr in the case of panel c) and up to 250 Myr
for panel b), and subsequently it accelerates again up to $\sim$ 2
Gyr. After this age a plateau in average Li abundance is present
in the case of the three oldest groups of clusters (i.e.~up to the
age of NGC~188), meaning that depletion might become
ineffective, as discussed by Randich
et al.~(\cite{R03}); note however that in panel c) only one star of
NGC~188 is present, with \nli~much higher than that of other old
clusters, thus it cannot be considered as significant in the
overall comparison. In these \teff~ranges the spread among stars in
M~67 is even larger than for stars considered in panel a): the
mean \nli~for the upper and lower envelopes of M~67 differ indeed
by a factor $\sim$ 8
for stars with $T_{\mathrm{eff}}=5900\pm150$ K,
and by a factor $\sim$ 4 at least,
for objects with $T_{\mathrm{eff}}=5600\pm100$ K. 
The Li content of the Sun is also shown in
Fig.~\ref{lit_age_nonstd}b and it is similar to that of the most Li-poor
objects in M~67 (and in the solar neighborhood; Pasquini et
al.~\cite{P94}).

Although stars in the three \teff~ranges investigated are
 characterized by a similar qualitative behavior
of Li depletion, it is important to note that
the quantitative amount of surface Li destroyed
within the various age intervals
increases for cooler stars.
The ``final'' abundances beyond 2 Gyr have indeed values of $\sim$ 2.5, 2.3
and 1.9 for stars in panel a), b), c) respectively
(and also the average abundance of the lower envelope in M~67
decreases --- 1.9, 1.3, and less). 

In summary, this discussion suggests that
late F and G-type stars within the three ranges
($T_{\mathrm{eff}}=6200\pm150$ K,
$T_{\mathrm{eff}}=5900\pm150$ K and
$T_{\mathrm{eff}}=5600\pm100$ K) have qualitatively similar
Li depletion histories which can be schemed as follows:
\begin{itemize}
\item a small amount of PMS Li destruction;
\item plateau during the early stages of the MS;
\item MS depletion;
\item plateau at old ages.
\end{itemize}
However, the cooler is the temperature, the higher
is the amount of Li destruction suffered during the various phases
and the shorter is the epoch of the early-MS plateau.

Two features are worth being stressed: first, Li depletion is
not a continuous process that can described with a $t^{-\alpha}$
law. Indeed Li depletion is characterized by different
timescales in the various age ranges. Second, 
the current homogeneous analysis of Li abundances in open clusters
covering a wide range of ages allows us to constrain for the first time
the age at which extra-mixing processes or other non-standard mechanisms
start working in solar-type stars and objects of similar temperatures.
Whereas solar-type stars do not undergo further
Li depletion during the early stages of the MS (convection
is not sufficient, since the convective zones --- CZ --- are
too shallow at these ages), an additional
mechanisms driving Li depletion appears around an age
of 150--250 Myr.

\subsection{Comparison with the models}\label{models_comp}

Several non-standard processes have been proposed in the last years
to be included in models:
mixing driven by rotation or by internal waves
(see below);
magnetic fields,
which might either enforce rigid rotation in stars 
or produce differential rotation (e.g., Charbonneau \& McGregor
\cite{ChMc92}; \cite{ChMc93}; \cite{ChMc96});
microscopic diffusion, determining the separation of atomic species in the stellar gas (e.g., Richer \& Michaud \cite{RM93}; Chaboyer et al.~\cite{CDP95a});
mass loss through stellar winds (e.g., Swenson \& Faulkner \cite{sf92}), causing Li dilution
in stars;
tachocline mixing (Piau et al.~\cite{piau}). The tachocline is a thin layer of separation between the CZ, in which differential rotation occurs, and the internal radiative core, rotating as a rigid body; tachocline mixing is a hydrodynamic process, not related to magnetic effects.

We focus here on those mechanisms 
for which detailed predictions of Li depletion for solar--type
stars as a function of
age and \teff~are available. Namely:
{\it (i)} slow mixing induced by rotational instabilities and angular
momentum loss (AML; Chaboyer et al.~\cite{CDP95a}, \cite{CDP95} --- 
hereafter CDP95b, dashed line in Fig.~\ref{lit_age_nonstd}; Deliyannis \&
Pinsonneault \cite{dp97} --- DP97, dot-dashed line/cross
in panel c)) and {\it (ii)} mixing driven
by internal waves
(Montalban \& Schatzman \cite{MS00} --- MS00, solid line). 
Note that the latter models include only gravitational waves,
neglecting rotational mixing. It is worth mentioning that
very recently other models have been developed which take into account 
the transport of angular momentum through
internal gravitational waves together with the effect of rotation
(Talon \& Charbonnel \cite{talon05}).

In the models by MS00 (see also
references therein) chemical mixing is induced by the
non-adiabatic propagation in the stellar interiors of internal
waves generated at the boundary of the CZ. 
Panel a) shows that these models
do not predict any surface
Li destruction
in stars hotter than 6000 K during
the whole age interval considered, at variance
with observations; on the other hand, too high Li abundances
are predicted for cooler stars (panel b))
up to an age of $\sim$ 3 Gyr. After this epoch
the depletion suddenly increases and the theoretical value for stars
with solar age
is only slightly lower than that observed for the upper envelope in M~67;
however, although models at older ages are not present, it is clear that
the plateau for objects older than 2 Gyr is not reproduced and that
the theoretical behavior is opposite to the empirical one.
The same happens for objects with \teff $=5600\pm100$ K (panel c),
but in this case the depletion predicted by the models
strongly increases around 1 Gyr, due to the deeper convective zones of these
stars. 
In conclusion, MS00 models are not able to
reproduce quantitatively and qualitatively
the observed time scales of Li depletion, thus
mixing induced only by gravity waves can be excluded as the
main process responsible for MS Li destruction.

Rotation can drive Li depletion since meridional circulation and/or
instabilities triggered by differential rotation in the stellar
structure induce a slow mixing over time scales comparable to
those of light element burning. The agent responsible for the
rising of differential rotation would
be an external torque, i.e.~AML (spin down): 
this process acts to generate radial gradients of angular velocity 
and instabilities leading to
AM redistribution and mixing of material. 
Stars born as fast
rotators can either undergo a large amount of AML during the PMS
or the early stages of the MS and lose a large amount of their
initial Li content, or they can continue to rotate rapidly
preserving a high \nli; initial slow rotators cannot suffer a
large amount of AML and as a consequence the stellar structure
adjusts itself on a circulation-free state. 

We report in the three panels of Fig.~\ref{lit_age_nonstd}
two different rotational models.
The models by DP97 include AML
and AM transport dominated
by the secular shear instability (Zahn \cite{zahn}),
while other parameters such as the velocity
estimates for meridional circulation
and other instabilities were not varied in that study (see also
Pinsonneault et al.~\cite{pinso89}, \cite{pinso90} for a detailed
description of the models).

The CDP95b models (see the quoted reference
and Chaboyer et al.~\cite{CDP95a})
include both AML and meridional circulation, treated as a diffusive process. 
In addition these authors take into account also
diffusive mechanisms not related to rotation (e.g., gravitational
settling).
Chaboyer and collaborators computed several models
by varying a series of parameters, namely:
the mixing length parameter; the He and metal contents ($Y$
and $Z$); the efficiency of overshooting; a constant multiplying
the diffusion coefficients ($f_{m}$); and five rotational parameters
including the initial rotational velocity and a critical
angular velocity at which AML reaches a saturation level.
We chose to represent the ``best fit'' model by CDP95b, i.e.
the one which in their opinion provides the best agreement with
observations of open clusters. This model is labeled as VN
in the reference paper; the initial rotational velocity
is 30 km s$^{-1}$.

Figure \ref{lit_age_nonstd} shows that in some cases it is
present a reasonable agreement
between the predictions of rotational models and the empirical patterns.
To be noticed that the models by CDP95b indeed predict a discontinuous
Li depletion. More specifically,
for stars with \teff $=6200\pm150$ K (panel a), the model
by CDP95b gives a rather good qualitative agreement with
the empirical behavior --- at least
up to 2 Gyr; whereas the  theoretical early MS plateau
is too high with respect to the empirical one (no PMS
depletion is predicted), as already stressed,
we are only interested in the differential amount of MS Li depletion,
thus the absolute Li value in the plateau is not important.
On the other hand, the slowing down of Li 
destruction after 2 Gyr is not reproduced adequately.
The DP97 isochrones are apparently not able to well fit
the shape of the early MS plateau, but actually models are available
for three ages only (100 Myr, 1.7 Gyr and 4 Gyr). Thus, data between
the early MS and the age of the Hyades 
are lacking for these models and we cannot draw any conclusion.
In any case, the DP97 model
 with $v_{\rm{rot}}=30$ km s$^{-1}$ is in strident 
quantitative disagreement with
the empirical evidence.

Similar arguments apply to stars in panel b) (\teff $=5900\pm150$ K)
and c) (\teff $=5600\pm100$ K). The two panels show that the value of
the early MS plateau is only slightly overestimated by the CDP95b models,
while a good agreement is present between the Hyades age and 2 Gyr. 
However, the plateau for old stars is not fitted, i.e. 
Li abundances decrease
towards the solar age. Also 
the models by DP97 
are not able to reproduce the shape of the late MS
plateau.
A too strong depletion is predicted by the 
model with $v_{\rm{rot}}=30$ km s$^{-1}$ and
\teff $=5900$ K, if compared
with open clusters; 
this track would be able instead to fit the solar abundance;
only one \nli~value (at the Pleiades age) is available for the track with
$v_{\rm{rot}}=30$ km s$^{-1}$ and
\teff $=5600$ K (represented by a cross in panel c)), 
much lower than the empirical one.

In summary, the approach of models
including slow mixing related to rotation (and diffusion)
appears reasonable since these models can somehow reproduce
the observed Li depletion pattern.
The CDP95b tracks are
able to reproduce the discontinuous rate of
MS Li depletion up to the Hyades
age, while no conclusion can be drawn for the DP97 models.

The models by CDP95b are probably best suited for
reproducing the early-MS plateau and the subsequent
acceleration of Li depletion, due to the competition
between rotation
and diffusion.
The theories by DP97 ignore microscopic diffusion,
which instead can have an important role in
particle transport: for example (see Chaboyer et
al.~\cite{CDP95a}) the diffusion of $^{4}$He induces
gradients in the mean molecular weight, which inhibit rotational mixing
(since extra energy would be needed to move material to zones of different molecular weight). On the opposite, rotational mixing might inhibit the separation 
processes of diffusion. In this sense, the plateau after the ZAMS
could be the result of this competition. Also AML and meridional circulation
might be competitive mechanisms; such combined effects are not present
in DP97 models. As final comments,
we remark 
that neither the CDP95b models nor those of
DP97 can reproduce
the plateau at old ages. 
In addition, these rotational models are not able to
reproduce the solar rotational profile measured by helioseismology.
In a very recent paper, Talon \& Charbonnel
(\cite{talon05}) presented models including gravitational waves and rotation,
showing that the combined effect of such mechanisms leads to
Li abundances for stars at the Hyades age in agreement with 
observations, and at the same time they are able to shape the solar
rotational profile.
\section{Conclusions}\label{lit_conclusions}

We presented a new homogeneous analysis of all the Li data
available in the literature for F, G and K-type stars in open
clusters. The need of a
dataset with all the clusters on the same scale of Li abundance
and \teff~arises from the fact that comparisons between samples
analyzed with different methods are affected by random and
systematic errors which cannot be estimated \emph{a priori}.
We have indeed shown that differences between 
the methods of analysis do exist.
The database will be made available on the web, in order to allow
homogeneous comparisons between the various samples
and to test updated models: it includes 21 open clusters
observed by various authors spanning the age range
5 Myr -- 8 Gyr and with metallicities $-$0.2$\lesssim$[Fe/H]$\lesssim$+0.2.

Based on the new Li abundances obtained by us for open clusters, we
have determined average Li abundances for stars in different temperature bins
and age ranges in order to investigate
the time scales of Li depletion
from ZAMS to late MS. 
For the first time we were able
to put a limit in age to the appearance of non-standard
mixing mechanisms at work on the MS: stars
do not appear to destroy Li in the early stages of MS, but Li
depletion starts again around an age of $\sim$ 150--250 Myr.
Then the destruction is
ineffective beyond $\sim$ 1--2 Gyr in late F and G stars,
except for a fraction of stars in M~67. 
Overall we showed that MS Li depletion is not a continuous process.
We also confirmed the existence of a plateau at old ages.
A \emph{caveat} to our results is represented by the fact that
ages and metallicities for the clusters have not been derived 
with a homogeneous method. Whereas our quantitative
estimates of time scales may be somewhat affected by this fact,
we believe that our qualitative results, and in particular, the existence
of a Li plateau at old ages do not change.

As to the possible extra-mixing mechanisms operating during the MS,
we compared the empirical scenario
with the predictions of models including
slow mixing related to gravity waves (Montalban \& Schatzmann
\cite{MS00}) and to rotation and angular momentum loss
(Chaboyer et al. \cite{CDP95}; Deliyannis \& Pinsonneault
\cite{dp97}). 

We found that gravity waves (at least when
not coupled with rotation --- models by MS00)
can be excluded as the main agent responsible for MS Li depletion,
while slow mixing induced by rotation might explain
to some extent the empirical behavior. In particular,
the models by CDP95b including AML, meridional circulation
and diffusion are those which appears at the moment
to be the most appropriate.
Nevertheless,
several uncertainties are still present and none of the
models proposed is able to reproduce the Li plateau
observed for solar-type stars in clusters older than the Hyades.
Understanding this empirical behavior is instead very important
for its connection with Primordial Nucleosynthesis and 
Galactic evolution.
We conclude that
improvements in theoretical work are certainly needed
for a proper comprehension of the properties of
mixing and Li evolution in stars.

\begin{acknowledgements}
We are grateful to D. Barrado y Navascu\'es, A. Ford, R.D. Jeffries
and D.R. Soderblom for sending us tables
with their Li data. We are indebted to S.N. Shore for helpful
discussion on this work. Finally, we thank the referee C. Charbonnel
for constructive comments on the manuscript.
\end{acknowledgements}

\section*{Appendix}\label{appendix_lit}

We report here a more detailed description of the computation
of random errors in Li abundances 
(due to uncertainties in \teff~and $EW$)
for the open clusters included in the database and presented
in Table~\ref{samples_lit}
(see also Sect.~\ref{random_lit}).

The clusters with published errors in $EW$s are the following:

\begin{description}
\item IC~2602: Randich et al.~(\cite{R01}) published the
$\Delta{T_{\rm eff}}$ for each star, but their temperatures were
derived in most cases from both $B-V$ (with
the S93a calibration) and $V-I$ colors,
thus in principle we cannot use directly these values. 
Since no error is quoted in the
$UBV$ photometry used (Prosser et al. \cite{prosser}),
we adopted a conservative $\Delta{T_{\rm
eff}}=\pm100$ K, as Randich et
al.~(\cite{R01}) did for stars with \teff~derived from only one color
index.
\item IC~2391: the same as for IC~2602
(Randich et al.~\cite{R01}). 
\item IC~2547: Jeffries et al.
(\cite{jef2547}) give errors in \teff~of $\pm$ 100 K, using $V-I$
colors from Naylor et al.~(\cite{naylor}). Errors in $B-V$ from
the same photometric source can be as large as 0.03, corresponding to
the same value
$\Delta{T_{\rm eff}}\sim\pm100$ K.
\item $\alpha$ Per: for the
sample of Randich et al. (\cite{R98}) we used the $\Delta{T_{\rm
eff}}$ quoted in the paper ($\sim\pm$ 100, 150 and 200 K),
since no error information is present in the source
of photometry (Prosser
\& Randich \cite{PR98}; Prosser et al.~\cite{prosser98}); 
Randich et al.~derived \teff~by averaging two different calibrations
(Bessel \cite{bessel}; Alonso et al.~\cite{alonso})
which provide temperature errors very similar to those 
obtained with the S93a calibration.
For
Balachandran et al. (\cite{bala88}; \cite{bala96}) see below.
\item NGC~2451: we used the $\Delta{T_{\rm
eff}}$ of $\pm$200 K quoted by H\"unsch et al.~(\cite{hun04}),
which derived the temperatures from $B-V$ colors with the S93a calibration.
\item Pleiades: (i) Garc\'{\i}a L\'opez et al. (\cite{GL94}): the
same as for $\alpha$ Per of Randich et al. \cite{R98} (but
Garc\'{\i}a L\'opez et al.~use three different
calibrations); (ii)
Jeffries (\cite{jefple}) does not compute neither Li abundances,
nor the temperatures (he only measures the $EW$s); we adopted
$\Delta{T_{\rm eff}}=\pm100$ K, which is a conservative value for
Pleiades stars. \item Blanco~1: we derived $\Delta{T_{\rm eff}}$
from errors in the photometry
(${0.01}\lesssim\Delta{B-V}\lesssim{0.03}$). \item NGC~2516: we
derived $\Delta{T_{\rm eff}}\sim70$ K from errors in the
photometry. \item M~35: we derived $\Delta{T_{\rm eff}}\sim50$ K
from errors in the photometry. \item NGC~6475: $\Delta$\nli~were
already derived by Sestito et al.~(\cite{sestito03}), also for the
samples of James \& Jeffries (\cite{JJ97}) and James et
al.~(\cite{J00}), using typical conservative $\Delta$\teff~of
$\pm$ 100 K. \item Hyades: we were able to compute the errors for
the two stars of Soderblom et al. (1990) for which we assumed
$\Delta$\teff $\sim50$ K ($\Delta(B-V)\sim0.01$). \item Coma Ber: in the three source
papers the authors quoted the uncertainties in the photometry; we
used these values to compute the errors in \teff~(ranging from 25
to 150 K). \item NGC~6633: conservative $\Delta$\teff~derived from
errors in the photometry are of $\sim\pm$ 100 K. \item NGC~752:
Sestito et al.~(\cite{sestito04}) computed errors for their sample
and for that of Hobbs \& Pilachowski (\cite{HP86}) assuming errors
in \teff~of $\pm$ 150 K (deriving from $\Delta(B-V)$,
see the discussion in the quoted
reference). \item IC~4651: $\Delta{T_{\rm eff}}=\pm50$ K from the
photometry, with the exception of two stars which have
$\Delta{T_{\rm eff}}=\pm100$ K. \item NGC~3680: we computed
$\Delta\log n(\rm Li)$ for the two stars in the sample of Randich
et al. (\cite{R00}) using their ${\Delta}EW$ and assuming $\Delta
T_{\rm eff}=\pm100$ K as in Pasquini et al. (\cite{P01}, see
below --- errors in colors are not quoted
in the photometry source by Nordstr\"om et al.~\cite{phot_nord}).
\item M~67: only Randich et al.~(\cite{R02})
give errors in $EW$s. For their data,
we considered $\Delta{T_{\rm eff}}=\pm50$ K,
as in Jones et al.~(\cite{jones99}; see below).
\item NGC~188: Randich et al.~(\cite{R03}) assumed
conservative uncertainties in \teff~of $\pm$ 100 K deriving from
the photometry.
\end{description}

These are instead the clusters for which we roughly
estimated an average
uncertainty in Li abundances, since errors in $EW$ are not quoted:

\begin{description}
\item NGC~2264: (i) Soderblom et al.~(\cite{sod2264}) quote only a
mean error in Li abundance, $\Delta\log n(\rm Li)=0.1$ dex; we
retained this value, since we used the same method of analysis.
(ii) King (\cite{king}) give uncertainties of
about 20--30 m\AA~in $EW$ and of $\pm$140 K in \teff~ for his
sample. From these values he obtained $\Delta\log n(\rm
Li)\sim0.25$, in agreement with us. 
Note that King used a different \teff~calibration (Bessel \cite{bessel})
however perfectly consistent with that of S93a.
\item IC~4665 (Mart\'{\i}n \&
Montes \cite{mm97}): $\Delta{EW}$=15 m\AA, $\Delta T_{\rm
eff}=\pm150$ K, using the
calibration of Arribas \& Mart\`{\i}nez Roger
(\cite{AMR88}); we found this calibration to give
errors in effective temperature very similar to ours.
Mart\'{\i}n \&
Montes quote an average final
error in Li $\Delta\log n(\rm Li)\sim0.3$, but we
find slightly lower values, $\Delta\log n(\rm
Li)\sim0.23$. 
\item $\alpha$ Per: Balachandran et al.
(\cite{bala88}; \cite{bala96}) quote $\Delta\log n(\rm Li)=\pm0.1$
deriving from uncertainties in $EW$s and $\Delta\log n(\rm
Li)=\pm0.2$ from uncertainties in \teff~of $\pm 200$ K. 
We cannot use this latter value, since, besides
the different calibration, we adopted a different
photometric source (Prosser
\& Randich \cite{PR98}; Prosser et al.~\cite{prosser98}), 
where errors in $B-V$ are not given; as to the error deriving from $EW$, we can trust
the value quoted by Balachandran et al.~since they
derived Li abundances with MOOG, which appears
to be consistent with the COGs of S93b (see Sect.~\ref{random_lit}).
Thus, we can compute indicative errors
by assuming $\Delta T_{\rm eff}\sim\pm100$ K (quoted
for most stars by Randich et al.~\cite{R98})
and $\Delta\log n(\rm Li)=\pm0.1$ deriving from $EW$ uncertainties:
we obtain errors in
\nli~ranging from $\sim$ 0.13 for the warmest stars up to $\sim$
0.17 for the latest spectral-type objects. 
\item Pleiades: (i)
S93b give $\Delta{EW}$=15 m\AA, without estimating the uncertainty
in \teff. Considering $\Delta T_{\rm eff}=\pm100$ K, which we
think to be conservative for this well studied cluster, we find
average $\Delta\log n(\rm Li)\sim0.13-0.17$ (the error increases
for the coolest stars). (ii) Jones et al. (\cite{jones96}):
$\Delta{EW}$=5 m\AA~for slow rotators and $\Delta{EW}$=20 m\AA~ for
fast rotators. Assuming $\Delta T_{\rm eff}=\pm100$ K we obtain a
conservative $\Delta\log n(\rm Li)$ of 0.15 dex, both for fast
rotators (which are mainly Li rich stars) and for slow rotators.
(iii) Butler et al.~(\cite{butler}): they quoted
an uncertainty of $\sim$ 15 \%~in $EW$; as for the \teff, we used
the errors quoted by Garc\'{\i}a L\'opez et al.~(\cite{GL94}), who
reanalyzed the Butler sample; errors in \nli~range from 0.15 to 0.40 dex.
\item M~34 (Jones et al. \cite{jones97}): $\Delta T_{\rm
eff}=\pm100-130$ K (using
the calibration of S93a) and relative $EW$ uncertainties of about 10\%.
It follows $\Delta\log n(\rm Li)\sim0.13-0.15$. 
\item Hyades: (i)
T93: $\Delta{EW}$=2 m\AA, $\Delta T_{\rm eff}\sim\pm50$ K,
from uncertainties in $B-V$ of $\sim$ 0.01
(conservative; see the detailed analysis in the original paper);
we find $\Delta\log n(\rm Li)=0.1$ for cool Li poor stars and
$\Delta\log n(\rm Li)=0.05$ for hotter stars. \item Praesepe
(Soderblom et al. \cite{S93c}): $\Delta{EW}$=5 m\AA, no
\teff~uncertainty quoted. Assuming $\Delta T_{\rm eff}=\pm100$ K,
errors in \nli~range from 0.1 dex for late F and G-type stars up
to 0.17 dex for later spectral types. \item NGC~3680: for stars of
Pasquini et al. (\cite{P01}), we started from $\Delta T_{\rm
eff}=\pm100$ K quoted by them and we assumed a conservative
uncertainty of 15\% in $EW$. Errors in \nli~turned out to be
$\sim$ 0.10--0.15 dex. 
Note that they used the calibration of Alonso et al.~(\cite{alonso}),
which results into errors similar to $\Delta T_{\rm eff}$ deriving
from S93a.
\item M~67: Jones et al.~(\cite{jones99})
quote average $\Delta{EW}$ of $\pm$10 m\AA~ and  $\Delta T_{\rm
eff}=\pm50$ K for his data (using
the S93a calibration), as well as for the previous samples
(see Table~\ref{samples_lit}) re-analyzed by them. 
These values lead to $\Delta\log
n(\rm Li)\sim0.06-0.08$ for stars with Li detection.
\end{description}

{}
\begin{figure*}
\psfig{figure=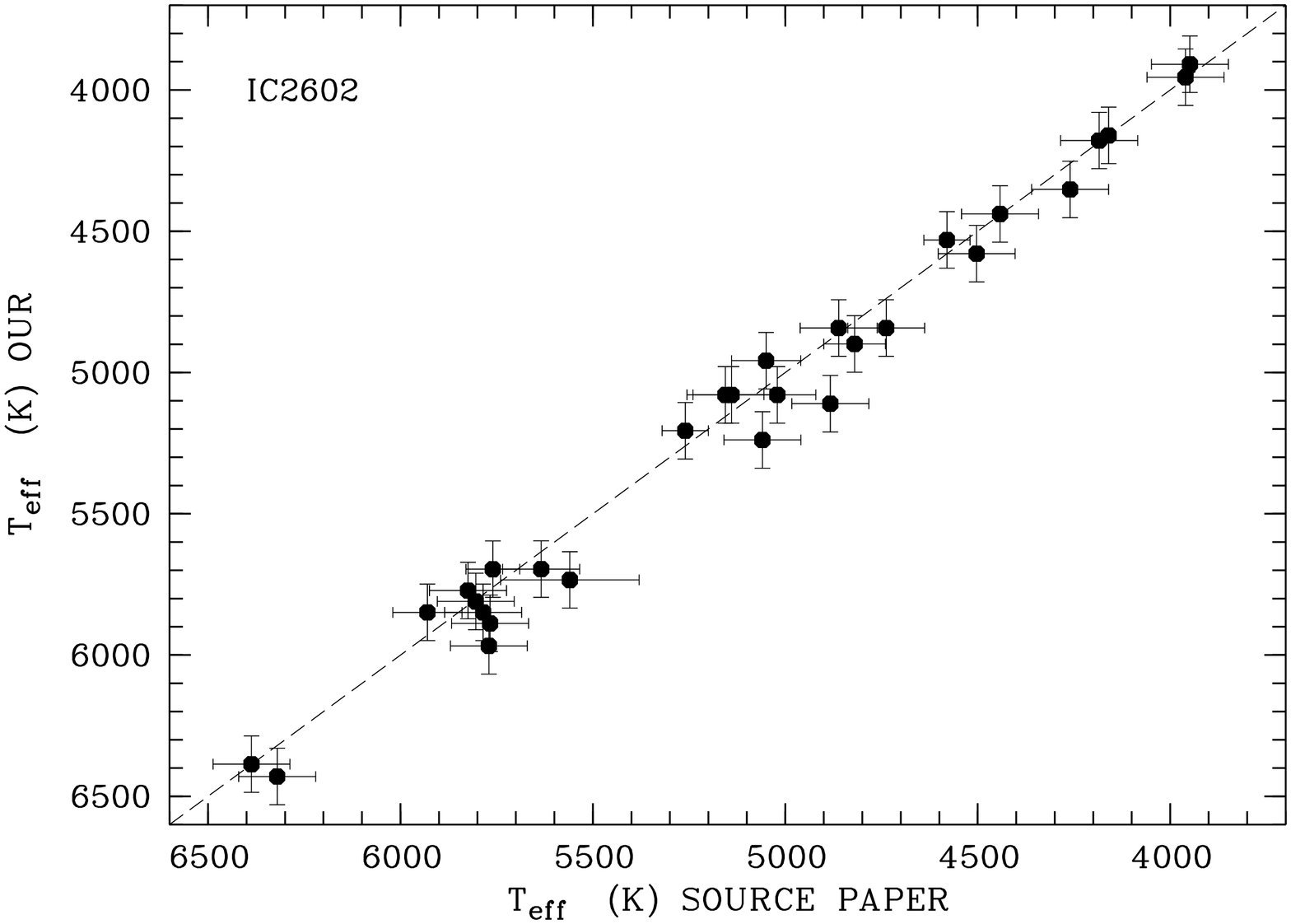, width=10cm, angle=-90}
\caption{Comparison between effective temperatures derived by us
using the calibration of S93a and those presented in the original
papers for IC~2602 (Randich et al.~\cite{R97}, \cite{R01}),
computed by
averaging two calibrations based on $B-V$ and $V-I$ colors.}\label{lit_Teff_cal_VI}
\end{figure*}
\begin{figure*}
\psfig{figure=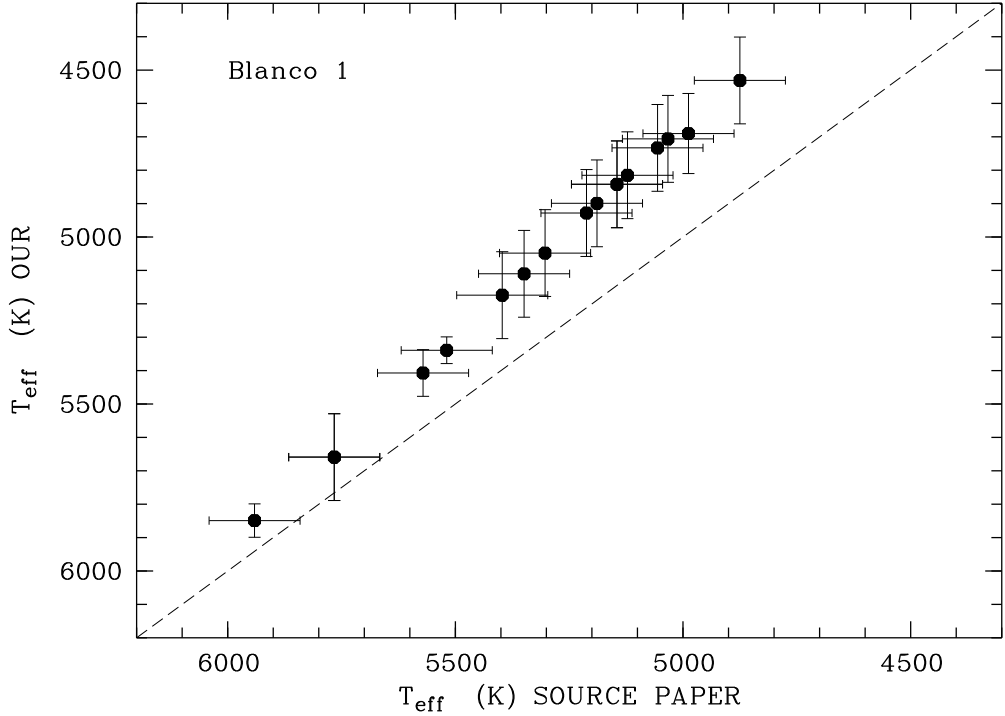, width=10cm, angle=-90}
\caption{Comparison between effective temperatures derived by us
using the calibration of S93a and those presented in the original
paper for Blanco~1 (Jeffries \& James
\cite{blanco1}), computed with the calibrations
of Saxner \& Hammarb\"ack (\cite{SH85})
and of B\"ohm-Vitense (\cite{bohm}).}\label{lit_Teff_cal}
\end{figure*}

\begin{figure*}
\psfig{figure=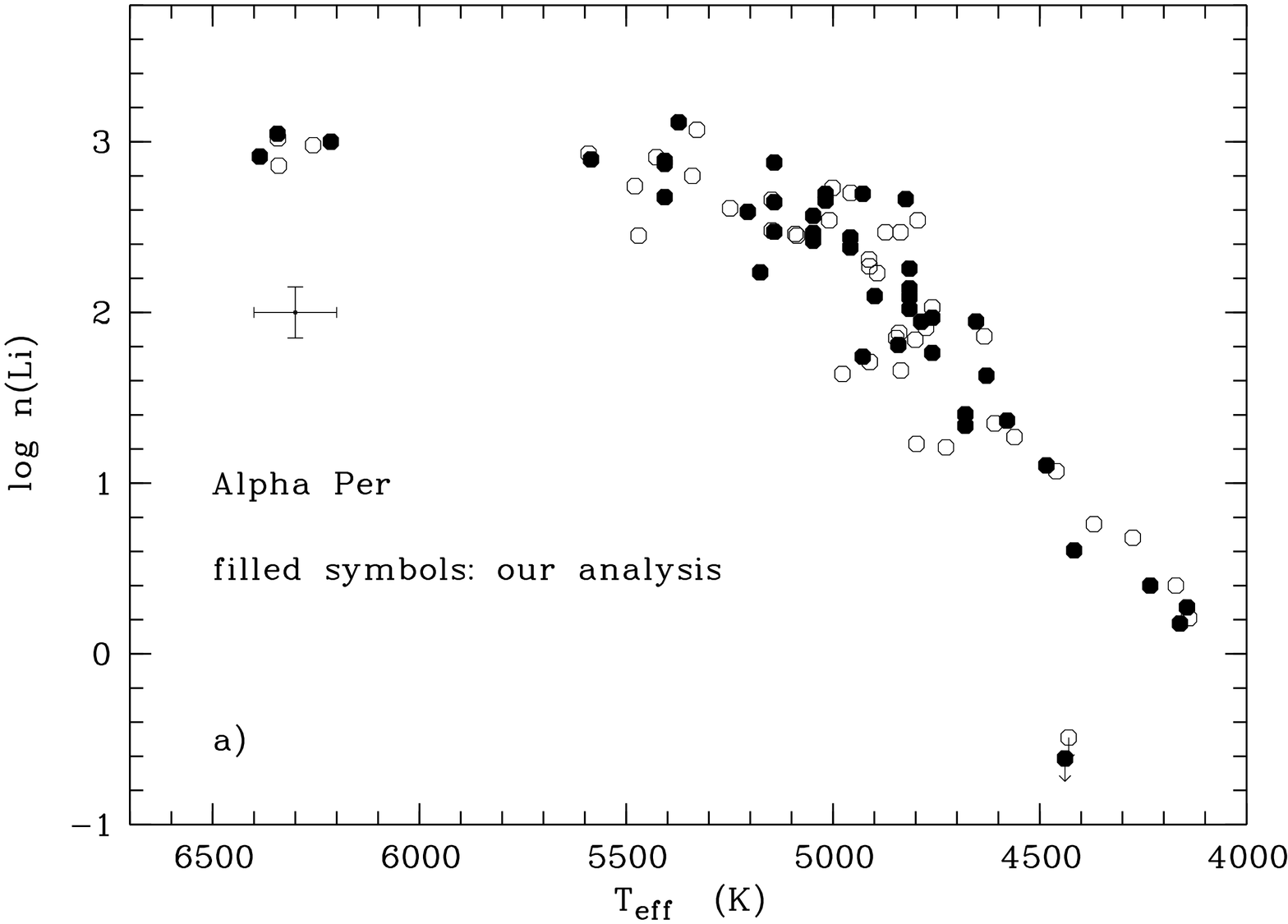, width=10cm, angle=-90}
\psfig{figure=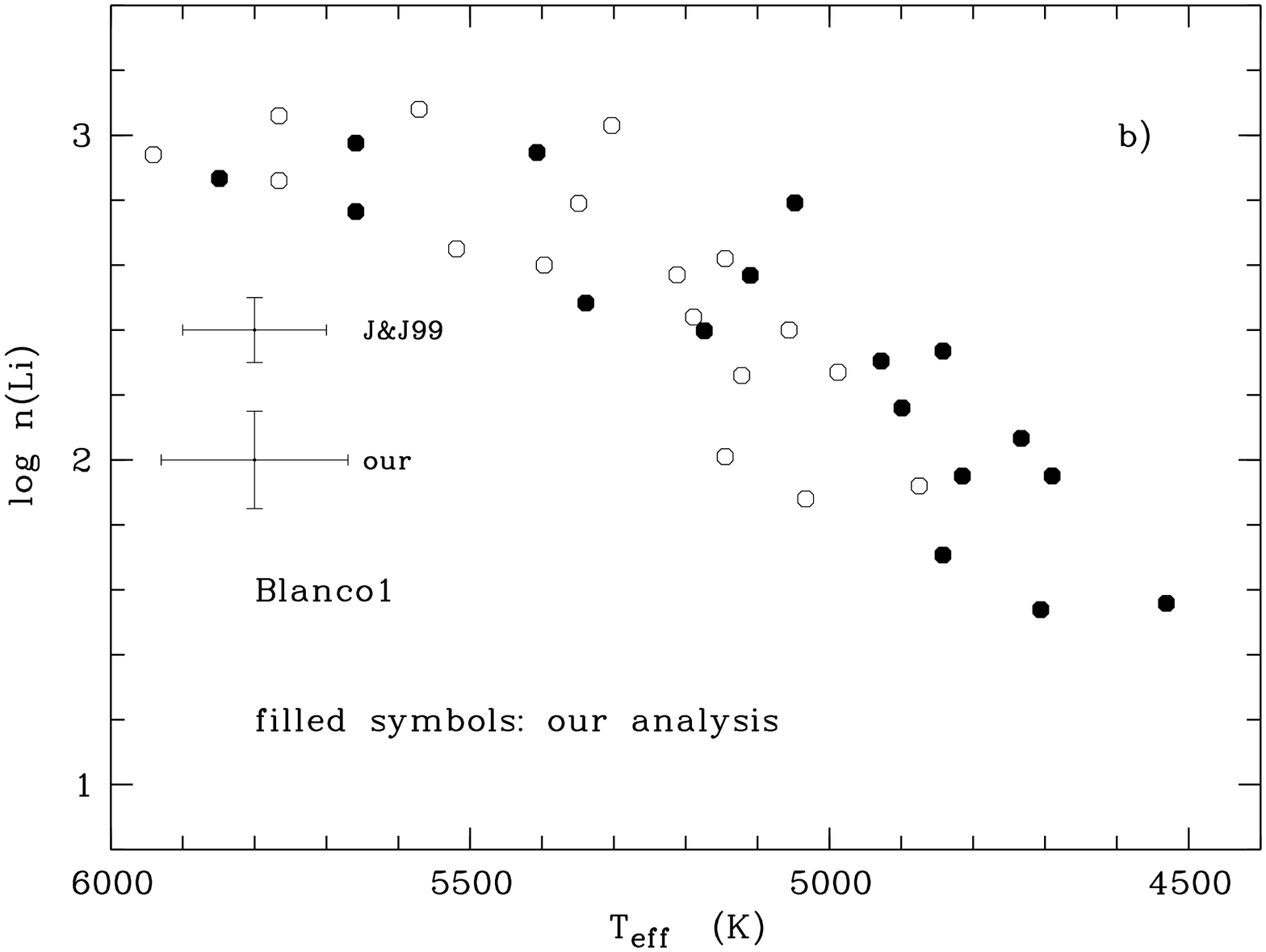, width=10cm, angle=-90} \caption{Li
vs.~\teff: the new distributions obtained with our method of
analysis are compared to the original data for two clusters ---(a)
$\alpha$ Persei (Randich et al. \cite{R98}) and (b) Blanco~1
(Jeffries \& James \cite{blanco1}).}\label{lit_comparison}
\end{figure*}

\begin{figure*}
\psfig{figure=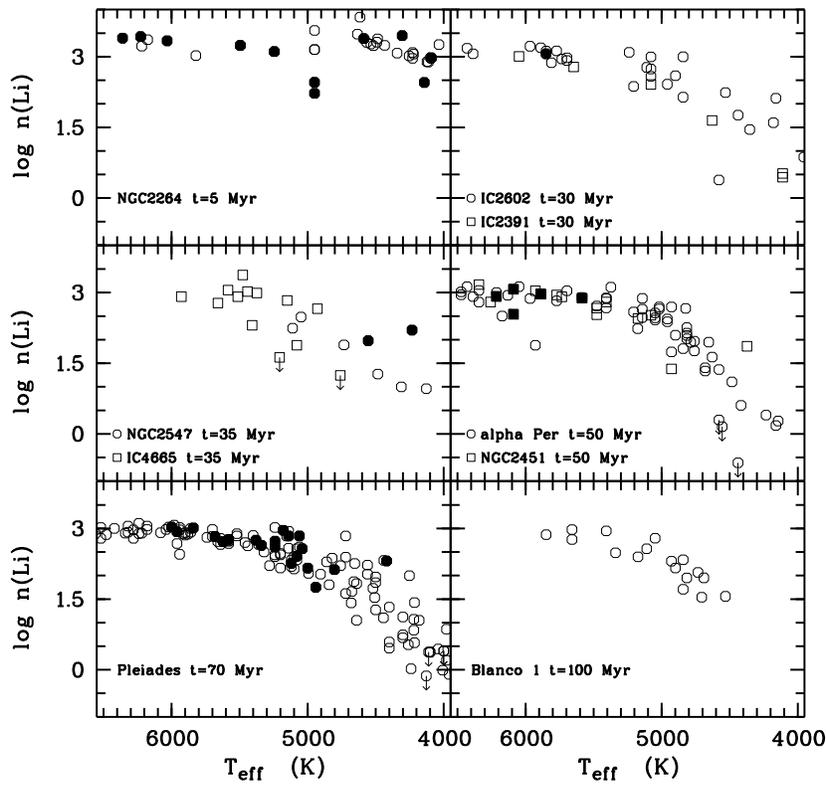, width=15cm, angle=0}
\caption{Li distributions for the open clusters in the database,
as derived by us: young clusters (age $\lesssim$ 100 Myr).
Filled symbols represent binary stars.}\label{OC_YOUNG}
\end{figure*}

\begin{figure*}
\psfig{figure=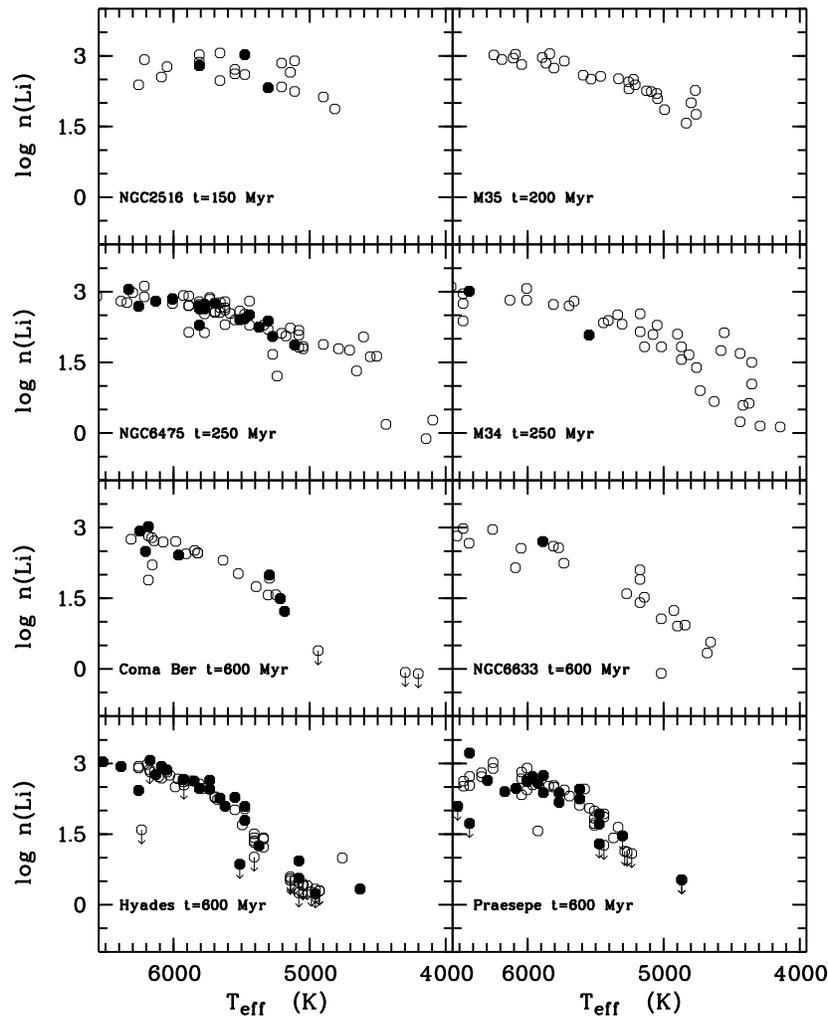, width=15cm, angle=0}
\caption{Li distributions for intermediate
age clusters ($\sim$150--600 Myr).}\label{OC_IM}
\end{figure*}

\begin{figure*}
\psfig{figure=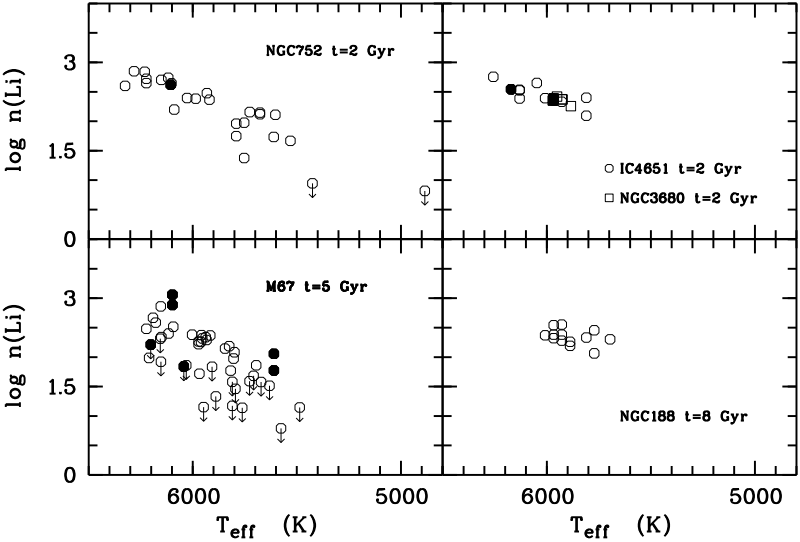, width=15cm, angle=0}
\caption{Li distributions for old open clusters
(age $\gesssim$1 Gyr).}\label{OC_OLD}
\end{figure*}

\begin{figure*}
\psfig{figure=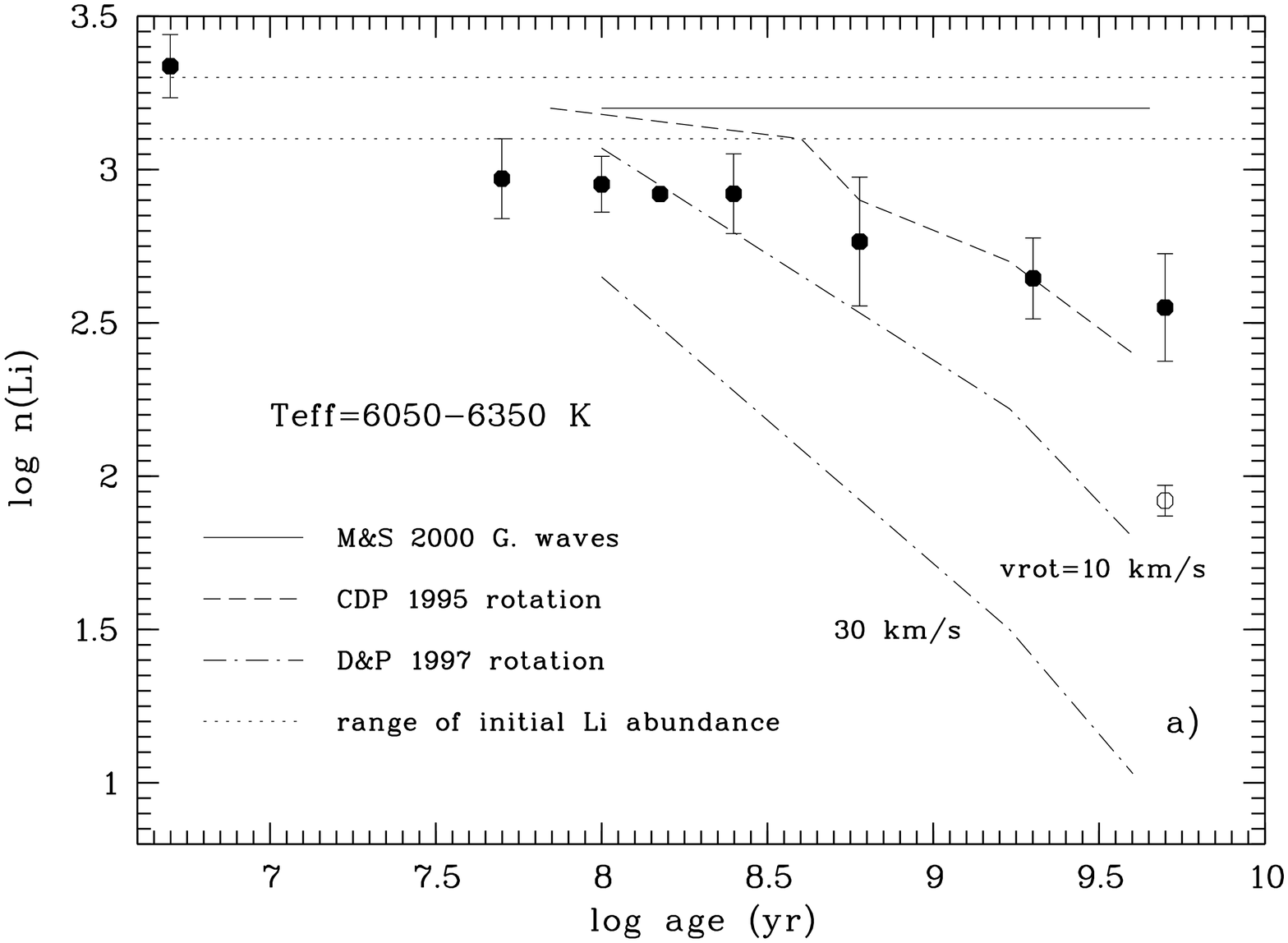, width=6cm, angle=-90}

\psfig{figure=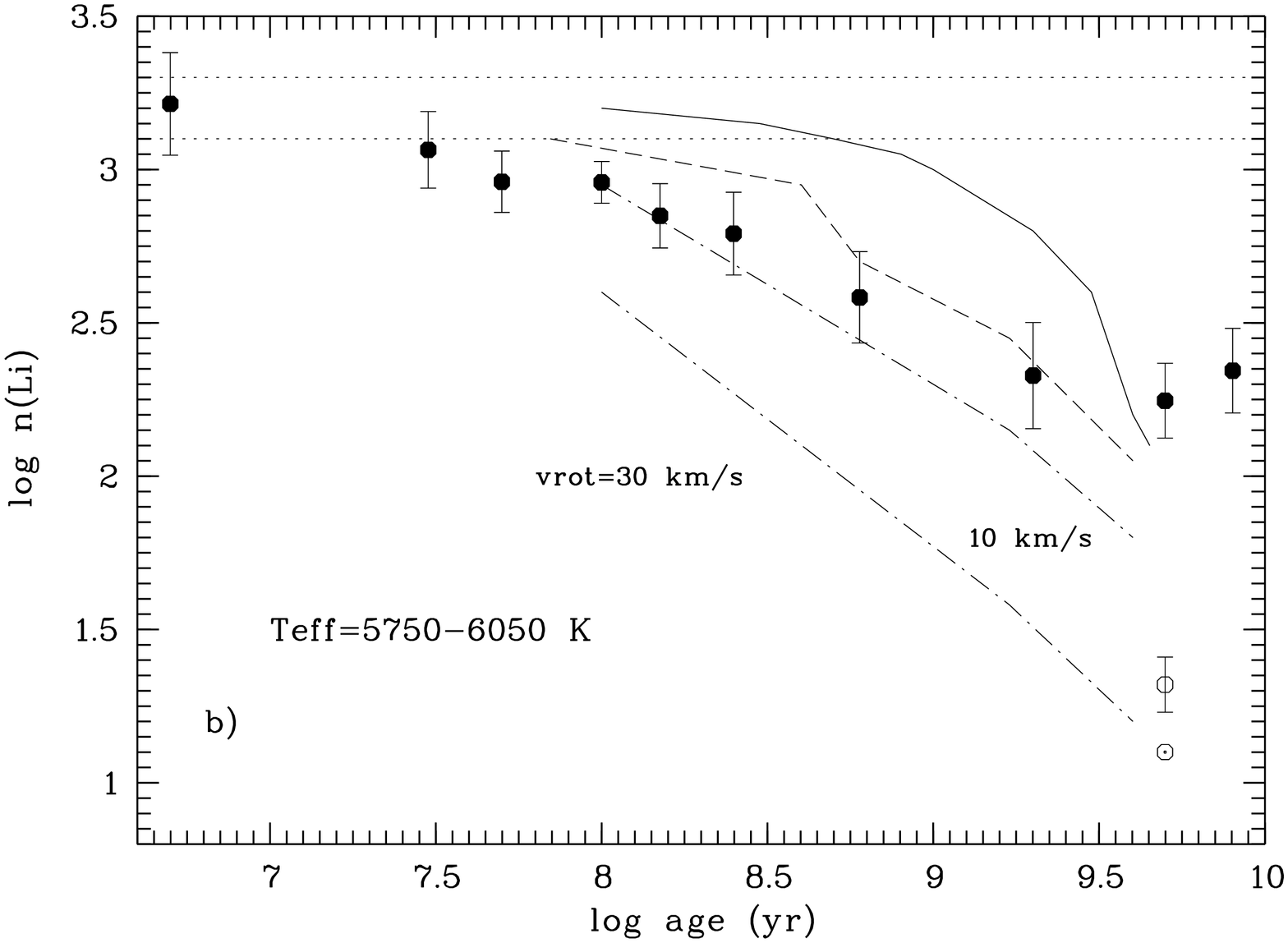, width=6cm, angle=-90}

\psfig{figure=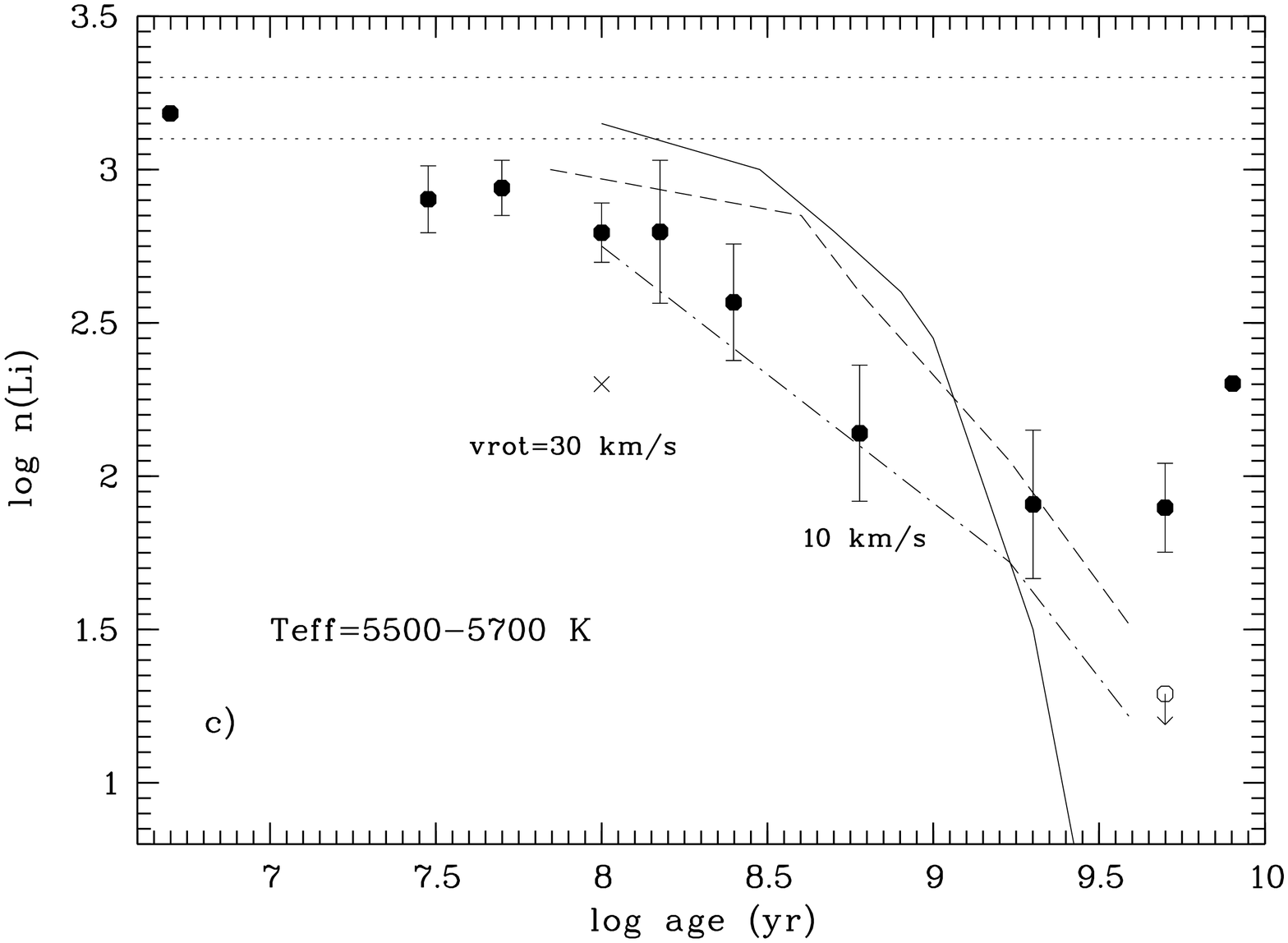, width=6cm, angle=-90}
\caption{Average \nli~as a function of age,
in three different \teff~ranges: a) $T_{\rm eff}=6200\pm150$ K; b) $T_{\rm eff}=5900\pm150$ K; c) $T_{\rm eff}=5600\pm100$ K. Clusters with
similar ages have been merged. The average abundance for the lower
envelope of M~67 (open
circle) and the Sun ($\odot$; panel b) 
are also shown. Dotted lines represent the range
of initial Li abundances for Pop.~{\sc i} stars.
Observations are compared to
predictions of non-standard models
 including
mixing by internal waves (solid line, Montalban \& Schatzmann
\cite{MS00}) and mixing induced by
rotation (dashed line, Chaboyer et al.
\cite{CDP95}; dot-dashed line, Deliyannis \&
Pinsonneault \cite{dp97}). The two models by  Deliyannis \&
Pinsonneault have different initial
rotational velocities.}\label{lit_age_nonstd}
\end{figure*}

\end{document}